# DeepFilters: Scattering-Aware Pupil Engineering with Learned Digital Filter Reconstruction for Extended Depth of Field Microscopy

Joseph L. Greene[1,2,*], Suet YIng Chan[1], Qilin Deng[1], Jeffrey Alido[1], Alexandra Lion[3,4], Guorong Hu[1], Ruipeng Guo[1], Tongyu Li[1], Kivilcim Kiliç[5,6], Ian Davison[3,5] and Lei Tian[1,5,6,*]

1 Boston University, Department of Electrical and Computer Engineering, Boston, MA, 02215
2 Current Address: Georgia Tech Research Institute, Electro-Optical Systems Lab, Atlanta, GA, 30332
3 Boston University, Department of Biology, Boston, MA, 02215
4 Harvard Medical School, Brigham and Women's Hospital, Department of Orthopedic Surgery, Boston, MA, 02215
5 Boston University, Neurophotonics Center, Boston, MA, 02215
6 Boston University, Department of Biomedical Engineering, Boston, MA, 02215

* Address all correspondence to Lei Tian. **Email:** leitian@bu.edu, Joseph Greene. **Email:** Joseph.Greene@gtri.gatech.edu

**Abstract:**

Extended depth of field microscopy encodes axial information into a single acquisition through engineered point spread functions, but conventional and deep optics approaches are subject to degradation in scattering tissue. We introduce DeepFilters, a scattering-aware deep optics framework that jointly optimizes a parameterized pupil filter and a digital-filter-based reconstruction network through a calibrated differentiable forward model to achieve broad generalization without retraining. Incorporating empirical scattering kernels, physics-guided regularization, and a hybrid genetic-gradient initialization strategy, DeepFilters extends the PSF from 16 µm to >400 µm in clear media and enables signal recovery beyond 120 µm deep in biological tissues, validated across fixed brain slices and sea urchin embryos.

## 1. Introduction

Fluorescence microscopy enables cellular-resolution imaging in 3D biological specimens, but increasing numerical aperture (NA) to achieve high spatial resolution reduces both the depth of field (DoF) and the field of view (FoV), limiting the accessible imaging volume [1]. Pupil engineering mitigates this tradeoff by redistributing how spatial and axial information are encoded, without sacrificing NA. By introducing controlled phase or amplitude modulation at the microscope's pupil, the system's 3D point spread function (PSF) can be tailored to extend depth coverage while preserving lateral resolution [2,3]. Engineered PSFs can exhibit axially varying profiles that encode depth information, enabling snapshot volumetric recovery from 2D measurements and facilitating high-speed 3D microscopy [4,5]. However, achieving high-fidelity 3D reconstruction over extended axial ranges remains challenging, as shaping a 3D PSF from a 2D pupil is inherently underdetermined, and recovering volumetric information from 2D measurements is an ill-posed inverse problem that typically requires computationally intensive reconstruction algorithms [6,7]. Extended depth of field (EDoF) techniques mitigate these challenges by utilizing a compact, slowly varying PSF over defocus to recover high-resolution lateral features at the cost of axial resolution [1,3,8-10]. In sparse imaging regimes, this tradeoff is beneficial as EDoF PSFs require minimal reconstruction to support high-throughput acquisition at reduced computational cost [10,11].

In practice, EDoF engineering faces several constraints. Engineered PSFs exhibit strong sidelobes, which introduce background artifacts and reduce contrast between neighboring emitters [10,12]. Additionally, many EDoF designs draw from a limited set of known phase profiles, such as axicons, cubic phase, and spherical aberration terms, which restricts the accessible design space during optimization [13-15]. These challenges become more pronounced in scattering specimens, where scattering-induced wavefront distortions degrade PSF fidelity, increase background, and limit depth penetration [16].

Recent end-to-end "deep optics" frameworks address some of these constraints by jointly optimizing parameterized optical elements and reconstruction algorithms through differentiable models [17-21]. In principle, this approach expands the accessible design space and enables tailored optical solutions. In practice, these pipelines often rely on simplified forward models coupled with large black-box reconstruction networks, introducing two key failure modes. First, the simplified forward models lead to mismatch between training and experiment, particularly when scattering is omitted or inaccurately modeled. Second, large reconstruction networks tend to overfit to training distributions, resulting in poor generalization to and across experimental data [17,21].

Additionally, mapping 2D pupil phase to 3D light distributions is inherently ill-posed, leading to unstable optimization that depends strongly on initialization and gradient quality [22]. Prior strategies, including aberration calibration, retraining on experimentally captured PSFs, alternating optimization schedules, and physics-inspired loss terms, partially address these challenges, but performance remains sensitive to initialization [17-19]. The black-box nature of large reconstruction networks further compounds these problems by obscuring learned inductive biases, making performance difficult to interpret and diagnose.

Here, we introduce DeepFilters, a scattering-aware deep optics framework that addresses these limitations through four key innovations. First, to balance forward-model fidelity with computational efficiency in scattering tissue, we introduce a calibrated proxy scattering kernel computed from a stochastic beam propagation model [23]. Second, to stabilize optical optimization, we develop a hybrid genetic-gradient initialization strategy that enables broad exploration of the pupil phase landscape prior to gradient-based refinement. Third, to explicitly control the EDoF profile, we introduce regularization strategies that enforce a prescribed axial extent of the PSF within specified depth ranges. Finally, to achieve robust and generalizable reconstruction, we depart from large black-box networks and instead introduce FilterNet, an unrolled sequence of differentiable filtering operations with physically interpretable parameters, while substantially reducing the number of learnable parameters.

Together, the DeepFilters framework mitigates sensitivity to forward-model mismatch, enables user-controlled EDoFs (demonstrated here up to 25x the nominal DoF), and promotes generalization across experimental conditions without retraining. We experimentally validate DeepFilters on both scattering phantoms and biological samples spanning a range of refractive indices and scattering conditions, recovering features across varying background levels and spatial scales. An overview of the DeepFilters computational pipeline and its experimental implementation is shown in **Fig. 1**.

These results establish DeepFilters as a principled and scalable framework for deep optics, demonstrating that physically grounded, interpretable models can achieve robust and generalizable performance without reliance on large black-box networks. More broadly, this work points toward a paradigm in computational imaging that tightly integrates optical design, physical modeling, and reconstruction, enabling controllable and high-throughput volumetric imaging in complex scattering environments.

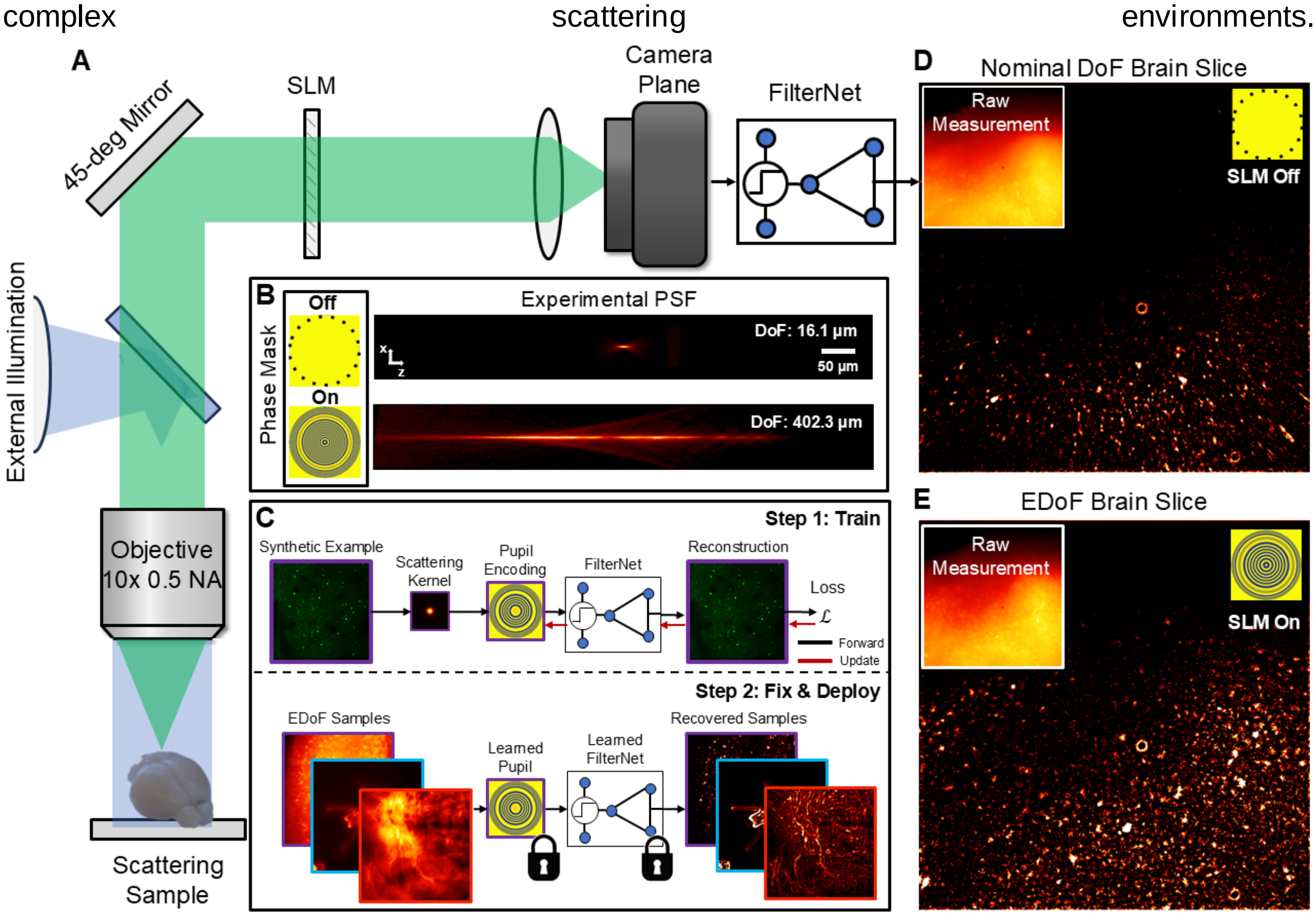

**Fig. 1. DeepFilters overview.** (A) Schematic of the experimental setup, where the EDoF phase profile is implemented using an SLM. (B) PSFs without (top) and with (bottom) an EDoF encoding for ~400 µm axial elongation. (C) DeepFilters pipeline: the model is trained once on synthetic data generated from a scattering-aware physical model, then deployed on experimental measurements, achieving robust generalization across diverse sample conditions. (D) FilterNet reconstruction of a 400 µm fixed brain slice without EDoF encoding, and (E) with EDoF optimized for maximal depth penetration and signal recovery.

## 2. Methods

DeepFilters integrates four key components: a calibrated scattering-aware forward model, an interpretable FilterNet for robust reconstruction, a hybrid genetic-gradient initialization strategy, and an end-to-end optimization framework integrating physics-informed loss and regularization terms, as shown in Fig. **2.**

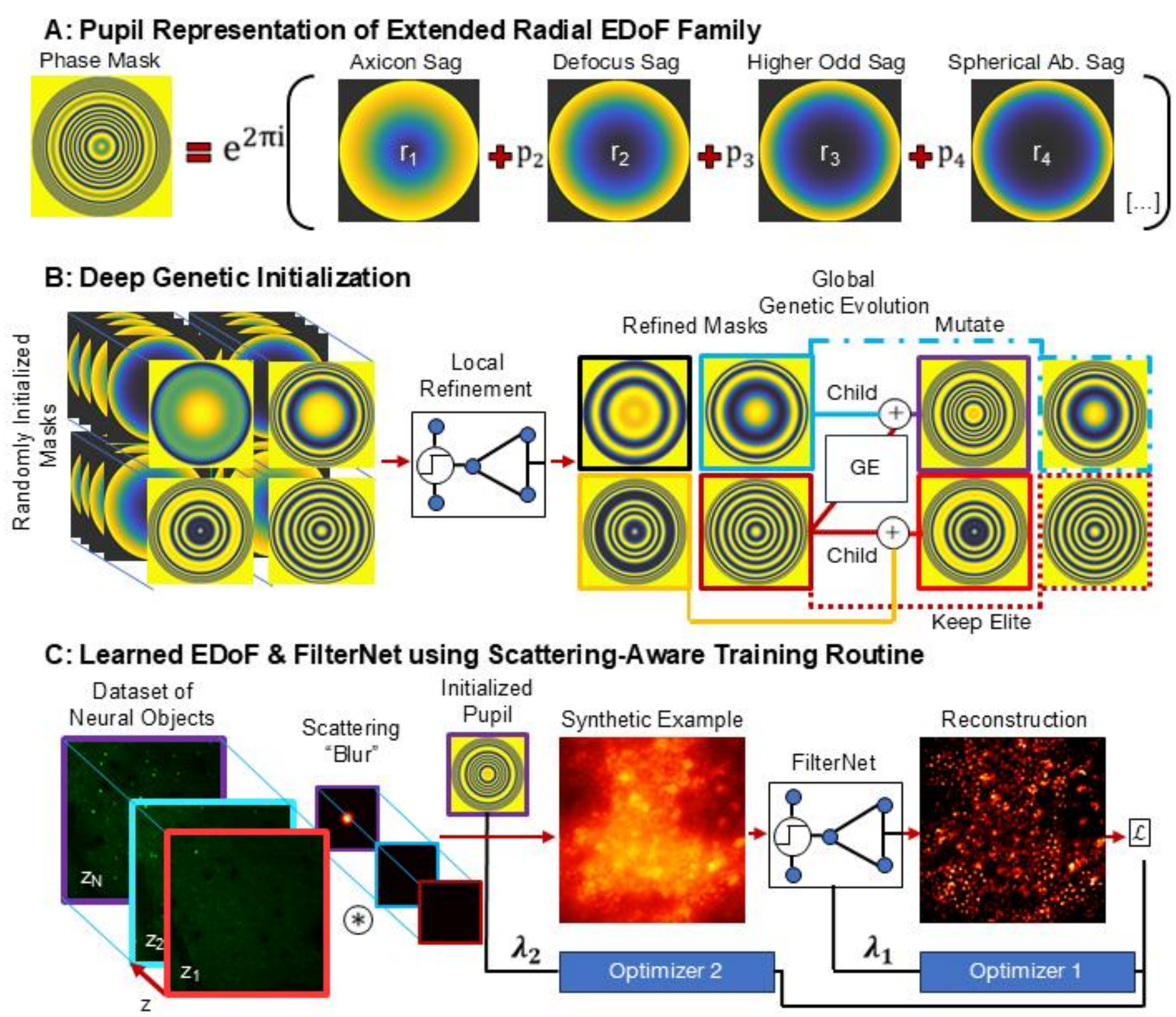

**Fig. 2. DeepFilters optimization pipeline.** (A) Pupil phase parameterized on an extended radial polynomial basis. (B) Deep genetic initialization combines global genetic search with local gradient refinement for robust starting conditions. (C) Joint optimization of the phase mask and FilterNet under a calibrated scattering-aware forward model.

*2.1 Scattering-aware forward model*

The simulation model follows the Fourier optics description of fluorescence imaging through a microscope with a pupil-plane phase modulation. An image formed from a 3D volume is

$$I(x,y) = \sum_{z\in Z} F^{-1}\{F\{o_I(x,y;z)\}\, OTF(u,v;z)\} \quad (1)$$

where $(x,y,z)$ are spatial dimensions, $(u,v)$ are lateral spatial frequencies, $o_I(x,y;z)$ is the discretized object fluorescence distribution, and $OTF(u,v;\, z)$ is the depth ($z$)-dependent optical transfer function (OTF), computed as the autocorrelation of the depth-wise coherent transfer function $P(u, v; z)$:

$$P(u, v; z) = M(u, v)\, D(u, v; z)\, A(u, v) \tag{2}$$

Here $M(u, v)$ is the learned pupil mask, $D(u, v; z)$ is the defocus transfer function, and $A(u, v)$ is the calibrated on-axis aberration, dominated by spherical aberration in our system. To limit the number of learnable parameters and to enforce radial symmetry, $M(u, v)$ is parameterized using a compact radial polynomial basis of order $j$:

$$M(u, v) = exp\,(i2\pi \sum_{j=1}^{J} a_j \rho^j) \tag{3}$$

where $a_j$ are the coefficients learned by DeepFilters in units of waves deviation and $\rho$ is the normalized radial pupil coordinate. Here, the pupil mask is optimized for a wavelength of $\lambda = 509$ nm. This basis captures classical EDoF phase functions, including axicon ($j = 1$), defocus ($j = 2$), and spherical aberration ($j = 4$), while allowing higher-order combinations that expand the accessible design space [10, 20] (**Fig. 2A**). Normalizing the radial pupil coordinate allows the same parameterization to be adapted across systems with different numerical apertures (NAs) (Supplementary Materials (**Fig. S13)**). We select six basis terms (i.e., $r^1$ to $r^6$) that capture commonly used radially symmetric EDoF phase profiles along with higher-order variations, while avoiding the convergence instability associated with over-parameterized representations, as characterized in the Supplementary Materials **(Fig. S2).**

To incorporate scattering, we use a calibrated proxy kernel $s(x, y, z)$ derived from a stochastic split-step beam propagation model, where random phase perturbations are introduced via layered random phase screens and parameterized to match the scattering mean free path $l_s$, anisotropy factor $g$, and mean refractive index of the medium $n$ [23]. This kernel is applied as a depth-dependent pre-blur to the object volume:

$$I(x, y) = \sum_{z \in Z} I^{-1}\{I\{o_I(x, y, z) * s(x, y, z)\} \times OTF(u, v, z)\} \tag{4}$$

In our model, we choose $l_s = 100\ \mu m$, $g = 0.9$, $n = 1.3$ to simulate scattering-induced PSF blurring from typical biological tissues. Using an averaged intensity kernel rather than full beam propagation is sufficient for incoherent fluorescence imaging and avoids the fine sampling constraints of split-step methods, enabling a fast and fully differentiable scattering model. This kernel calibration process and validation against split-step simulations are detailed in the Supplementary Materials **(Fig. S8)**.

*2.2* Interpretable FilterNet Reconstruction Module

FilterNet is implemented as an unrolled, differentiable version of the Dark Sectioning algorithm [24], augmented with adaptive median filtering to handle complex backgrounds and multiple noise sources, including shot, dark, and salt-and-pepper noise [25]. This approach combines selective background masking by identifying low-signal regions based on adaptive intensity thresholding and noise rejection by analyzing a locally sized neighborhood to remove impulsive noise while preserving sharp features. Unrolling allows the algorithm to be fine-tuned to training examples while keeping the functional operators interpretable, such that the reconstruction relies on physically meaningful mechanisms rather than purely learned correlations to improve robustness to variations across experiments. Together, these mechanisms remove diffuse background and outliers while maintaining underlying structural detail and generality. Only a small number of parameters are learned during training, specifically the background scaling factor, background removal factor, and low-pass cutoff for each Dark Sectioning iteration, while the overall architecture and operations are fixed by the underlying physical model [24]. The complete description of our FilterNet and its parameterization is found in the Supplementary Materials **(Fig. S5)**.

### *2.3* Deep Genetic Initialization for Reliable Convergence

To mitigate instability from randomly initialized optical parameters, we introduce deep genetic initialization, a hybrid initialization strategy that combines global genetic search with local gradient-based refinement. First, deep genetic initialization trains a surrogate FilterNet on the native PSF to learn high-contrast in-focus structures from a low-contrast defocused background. The surrogate is then frozen and used to evaluate a population of randomly generated pupil phase candidates. Genetic updates evolve the population toward lower-loss regions, while short gradient-based updates refine each candidate toward their local minimum. After several generations, this procedure yields a set of pupil masks with strong EDoF characteristics that serve as robust starting points for further optimization (**Fig. 2B**).

To quantify the contribution of each component of deep genetic initialization, we compare four initialization strategies on the same starting population: random sampling, random with genetic evolution, random with gradient refinement, and the full deep genetic approach. Random sampling alone achieves poor convergence and low EDoF extension due to the absence of iterative refinement. Adding a genetic step improves loss convergence but produces limited extension. Adding gradient refinement alone reduces loss but fails to converge the population. Only deep genetic initialization consistently achieves both convergence and meaningful EDoF extension, producing the lowest final loss across the tested population (Supplementary Materials, **Fig. S4**).

### *2.4 End-to-End Training Procedure*

The end-to-end training jointly optimizes the parameterized EDoF phase profile and FilterNet parameters. The initialized pupil candidate is co-optimized with an initialized FilterNet using simulated measurements generated by the scattering-aware forward model from synthetic volumes representing biologically relevant structures (**Fig. 2C**). After convergence, both the optimized phase mask and FilterNet parameters are fixed for deployment.

The training objective combines image-space supervision with PSF-level regularization. The image-space loss promotes the recovery of high-contrast in-focus features while suppressing low-contrast background. PSF regularization imposes explicit bounds on the EDoF target range by minimizing reconstruction error within the target depth, while maximizing reconstruction error outside the range to enforce rapid defocus of out-of-range structures. During optimization, this regularization is explicitly enforced by generating two images after the synthetic forward model within and outside the desired EDoF region, respectively. The image within the EDoF region uses the image-space objective function to minimize its loss to encourage sharp features in the optical PSF to minimize dissimilarity to a reference image (here, the axial projection over the EDoF region without optical blurring), while the image outside minimizes the negative of the objective to encourage defocusing of the optical PSF to maximize dissimilarity. Together, this approach enables direct control of the focal and defocus regions of the optimized PSF.

Additional constraints on peak energy and sidelobe strength further stabilize axial extension while maintaining lateral compactness. Without these PSF constraints, image-space losses alone produce ambiguous axial localization caused by depth compression in the image plane, as shown in the Supplementary Materials (**Fig. S1**). The full regularization formulation and gradient-balancing procedure are detailed in the Supplementary Materials.

To explicitly define the EDoF, the simulation volume is partitioned into a target depth range $Z_{EDoF}$ and its complement $Z_{Defo}$, specified by user-defined depth bounds, and optimized using:

$$L_{\square} = l(I_{EDoF}, GT_{EDoF}) - l\left(I_{defo}, GT_{defo}\right), \qquad 5$$

where $l$ compares the reconstruction $I$ with the corresponding ground truth projection $GT$ formed over each depth set (denoted by EDoF and defo) The objective $L$ minimizes the loss within $Z_{EDoF}$ while suppressing reconstruction outside this range, discouraging recoverable structure in $Z_{Defo}$. The scaling term $\alpha$ controls the rigidity of the EDoF region, with high $\alpha$ enforcing a strict cutoff and low $\alpha$ allowing some intensity leakage. Tuning $\alpha$ is important in setting the waist of the EDoF profile with respect to the EDoF cutoff, and in this work, we use $\alpha$ = 0.25 to achieve the results demonstrated here. This formulation encourages the learned PSF to produce high-contrast features within $Z_{EDoF}$ while suppressing by spreading the structures in $Z_{Defo}$ below the FilterNet detection threshold. Through this mechanism, the training routine can be explicitly parameterized to control the lateral and axial properties of the resultant EDoF profile.

The supervised loss $l$ combines mean squared error, structural similarity, Fourier-domain mean absolute error, and gradient-based mean absolute error to capture pixel-wise fidelity, perceptual quality, and high-frequency content in a joint objective function. PSF-level regularizers additionally constrain sidelobe energy and variation of the PSF peak axially to enforce desired physical characteristics independent of image-space feedback. Loss weights are set such that each term contributes approximately equal gradient magnitude at the pupil mask parameters, preventing any single term from dominating the update direction. The full regularizer formulations and their individual contributions to the learned PSF structures are provided in the Supplementary Materials (**Fig. S1**).

The pupil mask and FilterNet parameters are optimized simultaneously using independent learning rates and schedulers. The pupil mask is initialized using deep genetic initialization and is trained at $\lambda = 0.15$ with a super-convergence (e.g., one cycle) scheduler [26]. The superconvergence schedule was designed to ramp up to the desired learning rate over the first 10% of training, then decay to 1/1000 the value over the next 60% such that the pupil parameters have converged before training completes. FilterNet parameters are initialized from physically motivated values [24] and trained at $\lambda = 5 \times 10^{-4}$ without a scheduler. This training regimen was selected to encourage rapid exploration followed by convergence of the pupil mask, such that the FilterNet may fit to the learned EDoF without constant adaptation. These learning rates were selected such that every 1000 epochs corresponds with approximately 1 unit of change within each parameter (i.e., $a_j$ from **Eq. (3)**). DeepFilters training occurred for 10,000-20,000 epochs on an NVIDIA Tesla P100, V100, or Titan GPU on the Boston University Shared Computing Cluster.

*2.5 Training Dataset*

A custom dataset of 5000 high-resolution fluorescence images of neurons, vasculature, and cells across a range of scales and magnifications was assembled from the GenSat brain atlas, the Image Data Resource, and unpublished two-photon scans acquired by on-campus collaborators. High-NA source datasets were selected to avoid underestimating blur during synthetic image generation. All images were converted to 8-bit TIFF format and zero-padded to a uniform size. During training, images are randomly selected and stacked to form ground truth volumes, with 2D slices replicated axially to simulate objects with finite thickness beyond the axial resolution. Noise injection and background synthesis procedures used to match the power spectral density of real calibrated samples based on [27] and as described in the Supplementary Materials **(Fig. S10)**.

### 2.6 Experimental Setup

DeepFilters is deployed on a tabletop epifluorescence microscopy platform (**Fig. 1A**). This platform uses a 0.5 NA 10x objective (Nikon, CFI Super Fluor 10x) with an infinity corrected tube lens (Thorlabs, ITL200) to minimize aberrations. The platform contains two 4f optical relays to align a beamsplitter (Thorlabs, CCM1-BS026), linear polarizer (Thorlabs, LPVISE100-A), SLM (Meadowlark, E-19x12-400-800-HDM8), and camera (Imaging Source DMK 33UX546) on successive planes. To enable robust alignment, the SLM is mounted onto a 3-axis stage (Thorlabs PT3A). Samples are illuminated by an external LED source (Thorlabs, SOLIS-470C) and focused onto the microscope pupil through a high NA collimation lens (Thorlabs, ASL2520). Illumination and collection channels are decoupled through a green fluorescence protein (GFP) dichroic mirror (Thorlabs, MDF05-GFP).

## 3. Results

### 3.1 Experimental Demonstration of Controllable EDoF Up-to 25x extension

To validate the forward model and calibration procedure, we optimized EDoF phase profiles for target extensions ranging from 100 to 400 µm and measured the resulting PSFs by axially scanning 1 µm fluorescent beads with 2 µm steps using the corresponding phase pattern projected onto the SLM. Experimentally captured PSFs agree to within 12.8% across their optimized bounds, evaluated at the 1/e intensity cutoff with over-extension occurring more frequently than under-extension (**Fig. 3**). This level of agreement is achieved without explicitly requiring DeepFilters to cut off at the 1/e waist and only requiring a one-time calibration of the native spherical aberration in the system, whose conjugate is applied directly to the SLM pattern to correct model mismatch before deployment. All phase profiles were optimized from the same initialization produced by deep genetic initialization, reflecting the consistency of that strategy across target EDoF ranges and confirming that the calibration procedure does not meaningfully constrain the achievable design space. Here, the amount of extension was selected to match the thickness of the samples used during testing and is not limited by optimization. Reliable optimization up to 1 mm before destabilization, presenting the potential for interrogating samples thicker than this demonstrated limit, as shown in Supplementary Information, **Table 4**.

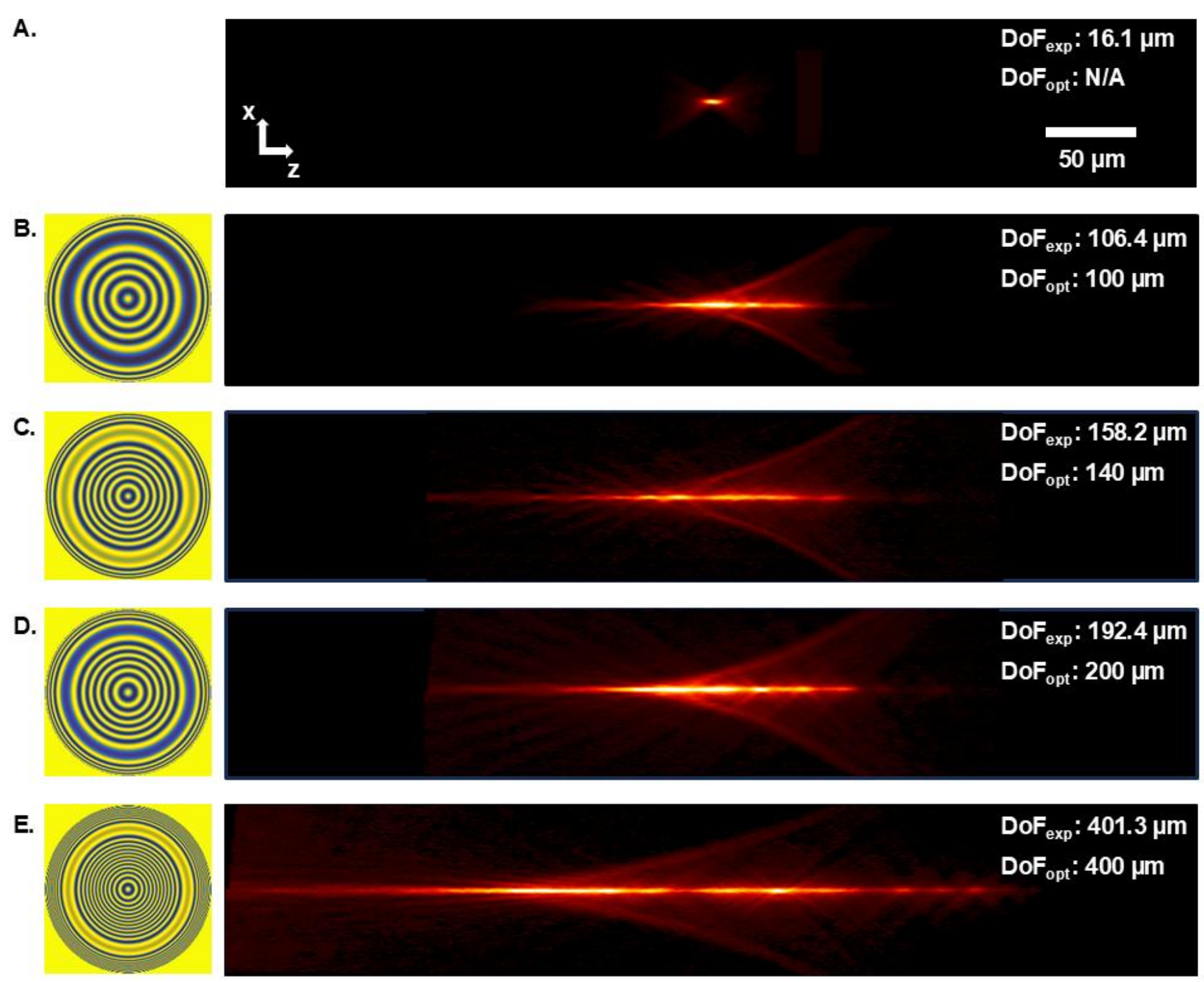

**Fig. 3. Demonstration of user-defined EDoF PSFs.** (A) Nominal PSF of tabletop platform. (B) Experimentally captured EDoF PSF for a designed EDoF region of 100 µm, (C) 140 µm, (D) 200 µm, (E) 400 µm

### *3.2 Robust Recovery Across Sample Morphology and Scattering Regimes*

To assess generalization across sample morphology and imaging conditions, we first imaged a fluorescently stained tilted tissue sample (Kimtech) to characterize the EDoF recovery of extended fiber structures in the presence of a spatially varying background. A depth-encoded maximum intensity projection (MIP) establishes the reference geometry of the tilted fibers, followed by snapshot images acquired without encoding, with a 200 µm EDoF, and with a 400 µm EDoF (**Fig. 4**). Without FilterNet processing, the spatially varying background obscures structural features across all PSF conditions. With FilterNet reconstruction, background is strongly suppressed, and fiber features are resolved across the full FoV and associated axial range (**Fig. 4B-D**). Notably, the FilterNet was originally optimized for neuron-sized objects and accommodating the larger fiber structures required adjusting only the patch size parameter governing local background estimation, reinforcing the intuitive adaptability of the reconstruction architecture across feature scales.

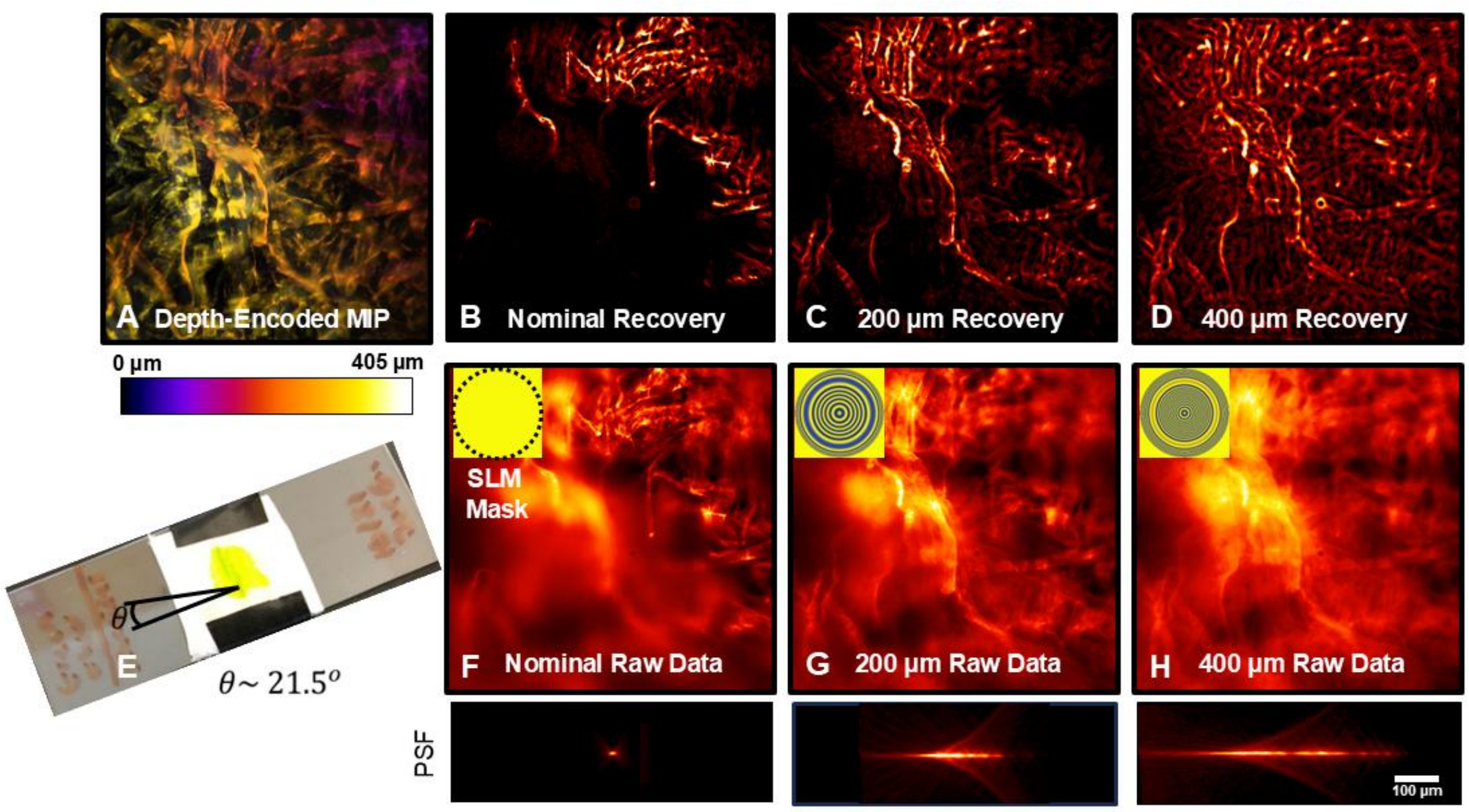

**Fig. 4. EDoF recovery of tilted tissue fibers.** (A) Depth-encoded MIP of a fluorescently stained fiber sample. (B-D) FilterNet-processed snapshots using the nominal PSF, 200 µm EDoF, and 400 µm EDoF, respectively. (E) Reference image of the fiber sample. (F-H) Corresponding raw snapshots without FilterNet processing, with SLM masks and PSFs.

We next validated performance on controlled bead phantoms across varying source densities, sample thicknesses, and scattering conditions to provide a more quantitative characterization of depth extension and scattering resilience. Phantoms were prepared from clear resin ($n = 1.51$, Formlabs) embedding 5 µm fluorescent beads as surrogate sources and 1 µm polystyrene beads for additional scatterers at controlled densities (Supplementary Materials, **Fig. S15).** Each scattering phantom was designed to exhibit an increasing scattering length (roughly 1 order of magnitude between each), and an EDoF was selected according to the final sample thickness. For each condition, a depth-encoded MIP acquired by axially scanning without pupil encoding serves as the reference, and FilterNet-processed EDoF snapshots are overlaid with the reference to assess spatial correspondence (**Fig. 5**).

In the lowest scattering phantom (e.g., $l_s = 22.5\ mm$), the 400 µm EDoF recovers sources down to 344 µm depth (**Fig. 5A**, purple ROI). At an increased source density of 12955 particles/mm$^3$, recovery remains feasible within the EDoF range, but background increases due to out-of-focus emitter accumulation, reducing contrast at the deepest recovered sources (**Fig. 5B**). In the highly scattering phantom with $l_s$ = 99 µm, signal recovery extends beyond one scattering length (**Fig. 5C**, blue ROI).

Three observations emerge across these conditions. First, the optimized EDoFs enable depth penetration across a variety of samples with variable scattering conditions, with access into a sample only being limited by scattering properties that encroach on the designed depth range (here observed to be slightly past one scattering length in the calibrated phantoms). Second, recovered EDoF signals exhibit less sample-induced aberration than the reference MIP, consistent with prior reports of aberration resilience in extended PSF designs [10], and further supported by stochastic modeling. Third, deeper sources also show progressive brightness reduction within the EDoF, which is attributed to reduced optical throughput as emitters move away from the focal plane. Practically, this throughput falloff represents a ceiling on achievable extension and becomes more pronounced as EDoF increases.

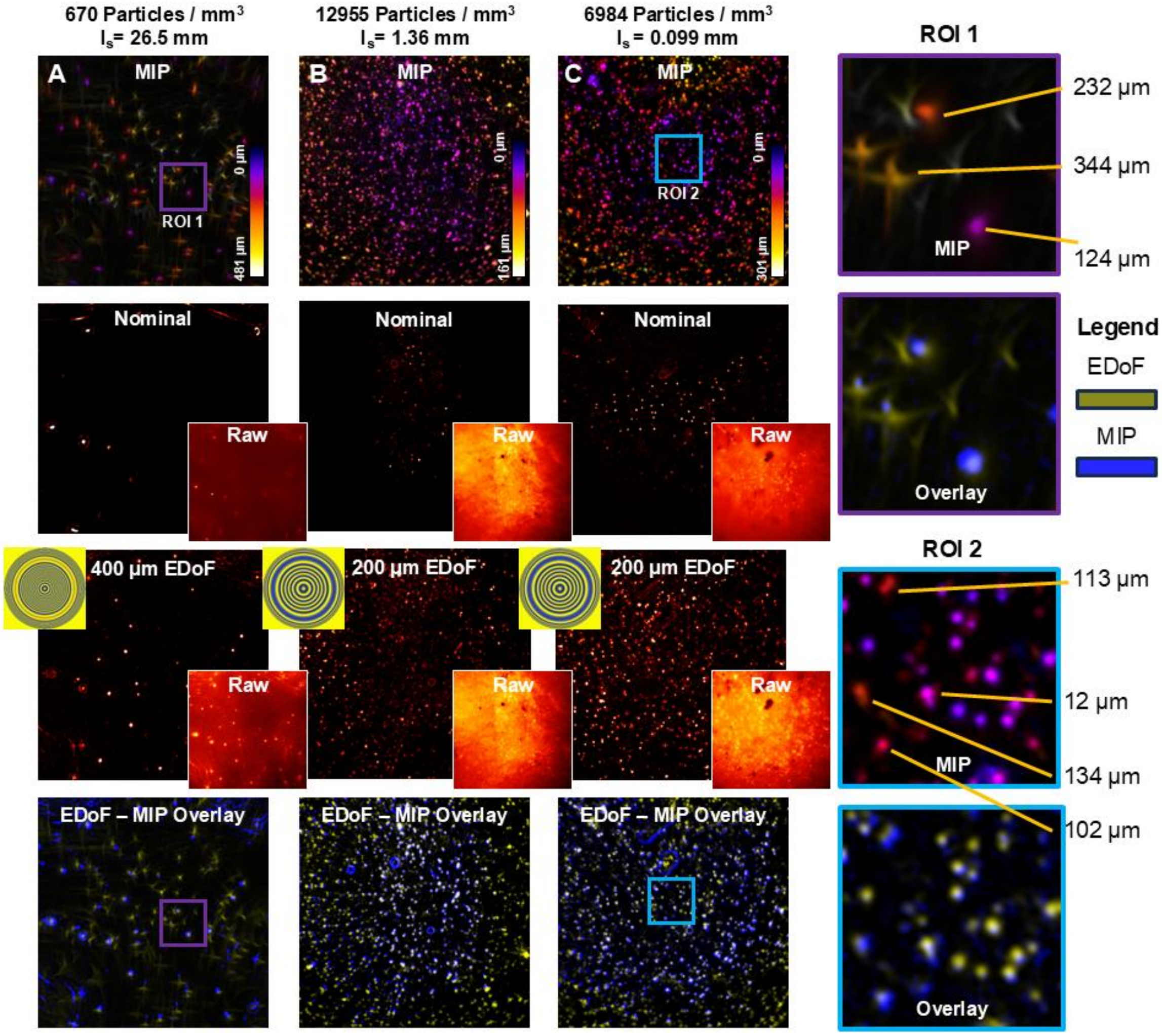

**Fig. 5. Controlled phantom validation across source density and scattering conditions.** (A) 480 µm phantom, source density 670 particles/mm³, 400 µm EDoF. Purple ROI indicates recovery at 344 µm depth.(B) 160 µm phantom, 12955 particles/mm³, 200 µm EDoF. (C) 300 µm highly scattering phantom with ls = 99 µm, 6984 particles/mm³, 200 µm EDoF. Blue ROI indicates recovery beyond one scattering length. Each panel shows the depth-encoded MIP reference (top), FilterNet recovery with nominal PSF (second), FilterNet recovery with EDoF PSF (third), and overlay of reference and EDoF recovery (bottom).

*3.3 Single-Shot Multi-scale Imaging of Sea Urchin Embryos*

Developmental studies of early morphogenesis require simultaneously resolving large-scale body patterning and cellular-scale structures across an embryo. We apply DeepFilters to sea urchin (*Lytechinus variegatus*) embryos 48 hours post-fertilization as a model for early morphogenesis and cellular migration [28]. Depth-encoded MIPs of captured embryos are shown alongside single-shot reconstructions without encoding and with a 200 µm EDoF PSF, with zoomed insets highlighting key internal structures (**Fig. 6**).

The 200 µm EDoF simultaneously recovers global structural metrics, including the body rod to post-oral arm rod ratio as a quantitative marker of larval health, and fine-scale features, including the blastocoel and primary mesenchyme cells, which are early indicators of blastula formation and skeletal development, respectively. Both the coronal and parasagittal orientations are shown to confirm structural recovery across imaging planes. The widefield architecture additionally supports simultaneous imaging of multiple embryos at 30 ms exposure times, enabling direct monitoring of developmental populations while resolving complete specimens (**Fig. 6A**).

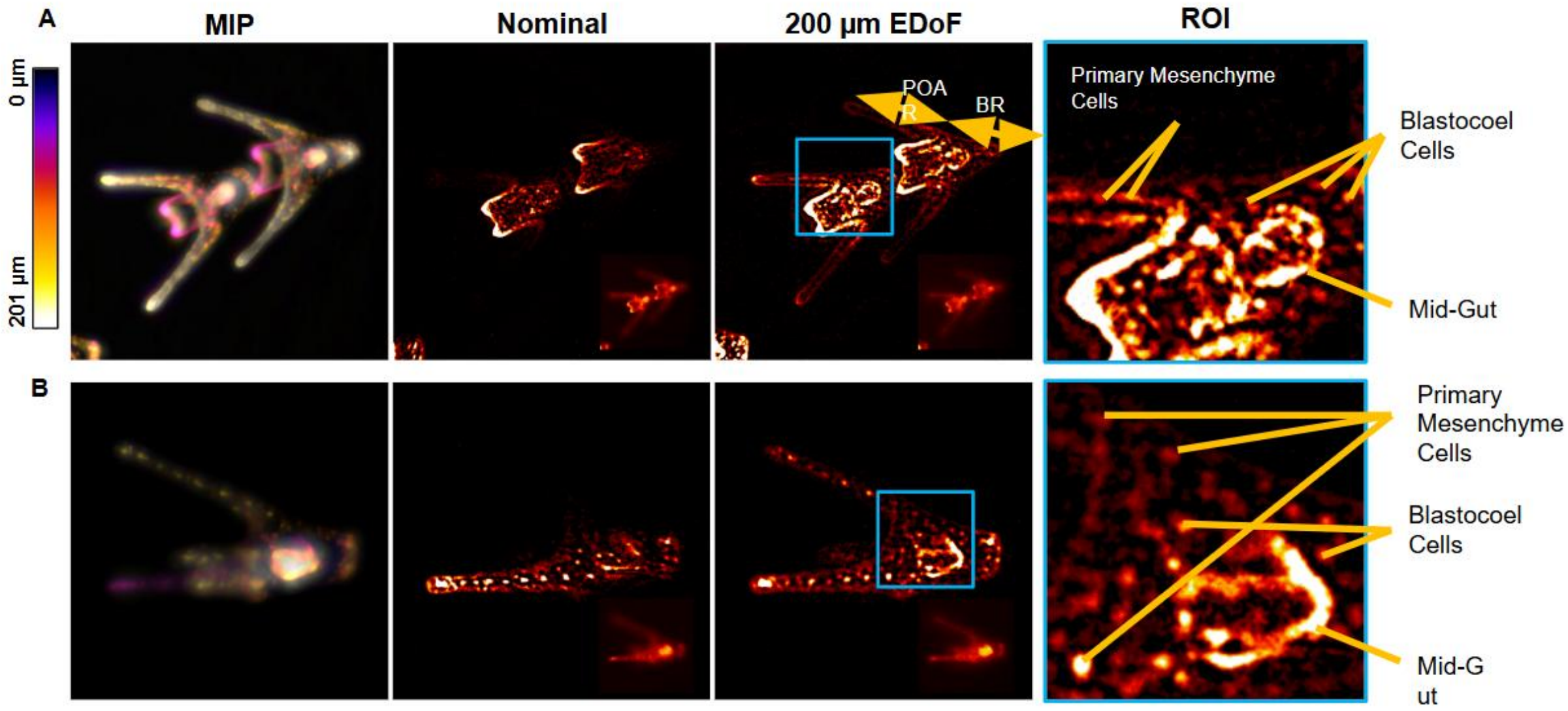

**Fig. 6. Single-shot global-to-local imaging of sea urchin embryos.** (A) Simultaneous imaging of multiple embryos at 30 ms exposure without and with 200 µm EDoF, compared to depth-encoded MIP. Blue ROI highlights fine internal structures. (B) Parasagittal plane reconstruction.

### *3.4 Neuronal Recovery Beyond Two Scattering Lengths in Fixed Brain Tissue*

Fixed mouse brain slices of varying thickness were imaged to characterize neuronal signal recovery under increasingly scattering conditions. Cortical tissue at green wavelengths exhibits a scattering length of approaching 100 µm [29], placing the full thickness range of tested slices between roughly one to four scattering lengths. The results for a 400µm thick sample are shown in **Fig. 7**, which compares the depth-encoded MIP, nominal PSF recovery, 140 µm EDoF recovery, and an EDoF-reference overlay to confirm recovered features. Additional data for 100µm and 200µm thick brain slices are included in (Supplementary Materials, **Fig. S15**).

Neuronal signals are recovered consistently beyond 120 µm depth across all slice thicknesses at NA = 0.5. A key finding is that deploying longer EDoFs beyond 140 µm does not extend the maximum recovery depth in brain tissue but does increase background due to enlarged PSF sidelobes, degrading contrast (see Supplementary Materials, **Fig. S8**). This behavior reflects a scattering-specific performance ceiling: once scattering-induced intensity loss dominates, additional axial encoding captures more background than signal. However, the ability to recover beyond one scattering length indicates utility in pushing this barrier by incorporating scattering properties during training.

To demonstrate the recovery of other complex morphologies in highly scattering media without tuning, DeepFilters is used to recover extended vascular reconstruction in a whole fixed mouse brain (Supplementary Materials**, Fig. S7**).

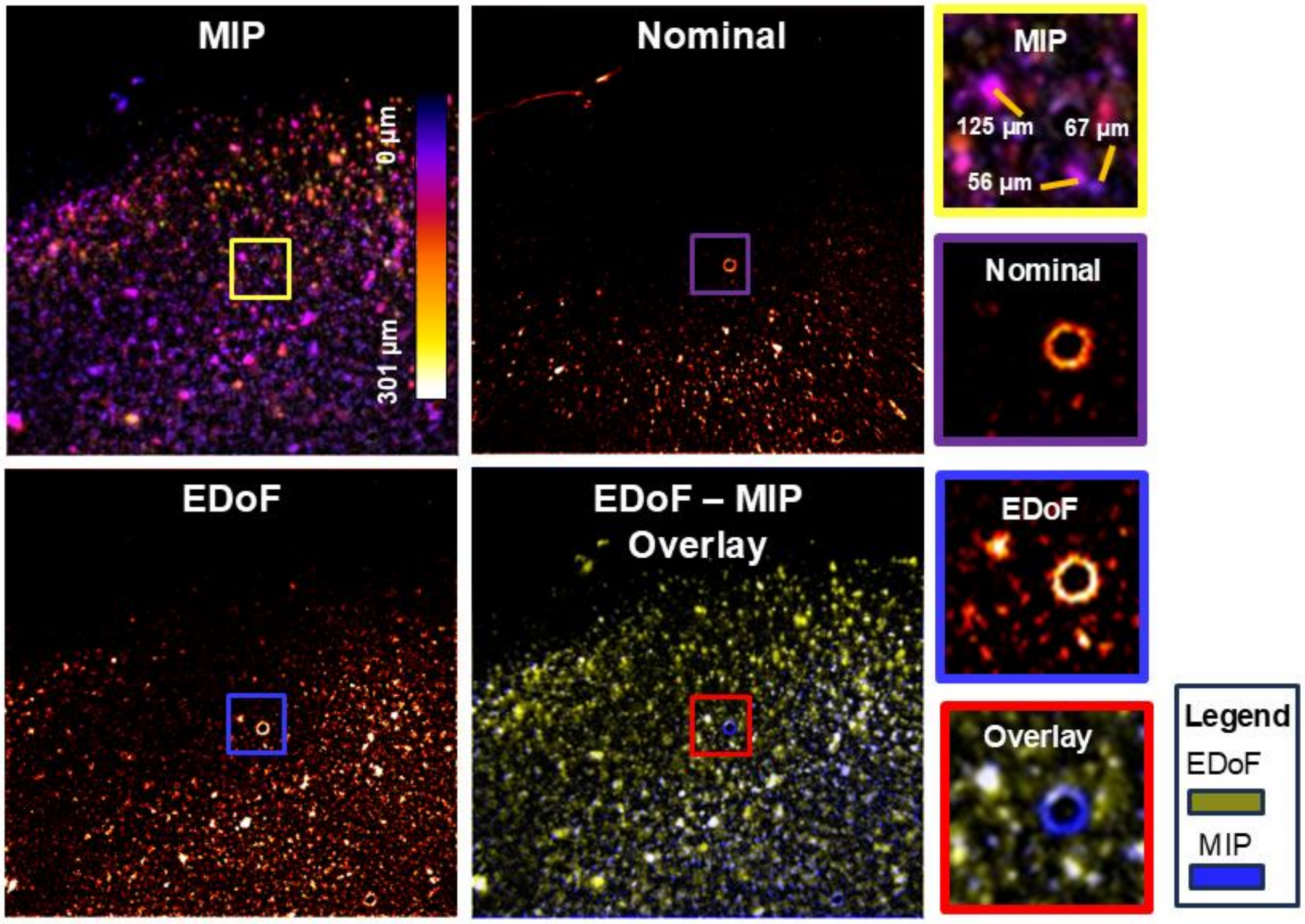

**Fig. 7. Neuronal recovery in a fixed mouse brain slice.** Comparison of depth-encoded MIP, nominal PSF, recovered PSF using 140 µm EDoF mask, and overlay for recovered EDoF (yellow) and MIP (blue) neurons for 400 µm bran slice with corresponding zoomed in ROIs indicating the deepest recovered neuronal signal.

*3.5 Robustness of DeepFilters Across Samples and Imaging Conditions*

A central claim of the DeepFilters framework is that a single optimized PSF and FilterNet, trained entirely on synthetic data, generalizes across the experimental conditions without retraining. The results spanning tissue fibers, controlled phantoms, sea urchin embryos, and fixed brain slices of varying thickness and scattering length were all obtained using the same set of trained parameters with no sample-specific retraining or architectural modification. The only adjustment made across experiments was the FilterNet patch size parameter, which governs local background estimation and scales intuitively with the dominant feature size of the sample.

Another key functionality of DeepFilters is the flexibility in the user-defined EDoF range. To quantify the robustness of this training process, we repeated end-to-end optimization across 10 independent random seeds for five target EDoF ranges from 200 to 1000 µm to demonstrate the low variability in the learned EDOF extensions. Learned phase mask coefficients showed standard deviations below 0.23 wavelengths even at the 1000 µm EDoF, and the achieved extensions track closely across all conditions **(**Supplementary Materials**, Tables S3-S4**). A sensitivity analysis confirmed that parameter variance of this magnitude produces no measurable degradation in PSF extension or reconstruction quality (Supplementary Materials**, Fig. S12**).

Taken together, these results support the conclusion that DeepFilters converges to consistent solutions under appropriate initialization and that the trained system transfers reliably to experimental conditions.

## 4. Discussion

DeepFilters provides a stable, interpretable, and scattering-aware framework for EDoF microscopy that extends the utility of deep optics into biological tissue. By combining a calibrated scattering proxy model, physics-guided PSF regularization, genetic-inspired initialization, and a differentiable filter-based reconstruction network, the framework achieves a controllable depth extension demonstrated up to 25x its native DoF in clear media and signal recovery beyond 120 µm in highly scattering samples, validated without retraining across brain slices, sea urchin embryos, fluorescent phantoms, and fiber samples. The interpretable design of FilterNets provides a generalizable alternative to black-box reconstruction networks in deep optics pipelines, with implications for any application where synthetic-to-experimental domain shift limits deployment performance.

In scattering media, we observe the ability to recover signals past one scattering length, however this extension is finite. In fixed brain slices, extending the optimized EDoF beyond 120 µm does not increase the maximum recoverable depth but does increase background due to enlarged PSF sidelobes, degrading contrast at the depths already accessible with shorter extensions. This behavior reflects a fundamental tradeoff between axial encoding and scattering resilience: longer EDoFs require more aggressive wavefront shaping, which produces stronger sidelobes that accumulate scattering-induced background more rapidly than they extend the usable signal. The scattering-aware forward model incorporated during training mitigates this effect by biasing the optimization toward profiles that balance extension with sidelobe suppression, as confirmed by the simulated comparison showing only 7.8% elongation loss in our optimized PSF under scattering versus 17-25% for conventional designs under fixed conditions (Supplementary Materials **(Fig. S8)**.

However, it is observed that the calibrated scattering kernel over-approximates scattering when compared to a split-step simulation (Supplementary Materials **(Fig. S8).** Developing a self-consistent calibration procedure that accounts for the encoding-dependent scattering response would reduce this approximation error and potentially extend the achievable recovery depth in tissue. Nevertheless, the results suggest that for strongly scattering tissues, there exists an optimal EDoF range beyond which additional extension is counterproductive, and that this range is dependent on tissue-specific properties.

A related observation from the phantom experiments is the observed reduction in brightness of deeper sources within the EDoF, even in weakly scattering samples (e.g., $l_s = 26.5\,mm$). This falloff arises from reduced optical throughput as emitters move away from the focal plane and may represent a practical ceiling on achievable extension that is distinct from the resolution or contrast limits typically used to characterize EDoF performance [3]. Characterizing this limit across NAs and EDoF range serves as a next step for extending this framework toward more extreme extensions.

The generalization of FilterNets across the biological specimens shown here follows from promoting a more generalized filter-based network. Whereas large reconstruction networks learn to invert a specific forward model and fail when the experimental PSF departs from its synthetic training counterpart, FilterNets instead rely on physically motivated filtering operations whose parameters have direct interpretable meaning, making them robust to practical sources of PSF and environmental (e.g., scattering, background) mismatch that is unavoidable in practice. In a quantitative benchmark against the U-Net reconstruction in the Supplementary Materials **(Fig. S11)**, FilterNets outperform the U-Net on experimental data despite being trained on the same synthetic distribution, indicating that the improved generality is not a function of training data quality.

DeepFilters requires no modification to the imaging optical path beyond placing a phase element at the pupil plane, which can be implemented using an SLM for real-time switching between optimized EDoFs or using a static diffractive element or metasurface for compact, low-cost deployment [30]. The FilterNet runs at 45 FPS on a NVIDIA RTX4070 Laptop for 512x512 images, supporting video-rate processing without dedicated hardware. The same parameterized optimization framework generalizes across imaging systems with different NAs, as demonstrated for 0.2 and 0.75 NA objectives in the Supplementary Materials **(Fig. S13)**, suggesting a straightforward path toward adapting DeepFilters to other platforms, such as miniaturized platforms for in vivo neural imaging [1,10] or multi-view configurations [31]. Additionally, the emergence of differentiable physical models across new technology, such as event cameras [32-34], and alternative applications, such as vision systems [35], creates avenues to extend this work to new applications. Future efforts may extend the parameterization used here to support full 3D PSF engineering or multi-focus designs [4-6, 31] to bridge this advancement to full volumetric reconstruction.

### Funding.

This project was funded by National Institutes of Health (R01NS126596) and a grant from 5022 - Chan Zuckerberg Initiative DAF, an advised fund of Silicon Valley Community Foundation. Joseph Greene acknowledges the National Science Foundation Neurophotonics Research Traineeship Program, Understanding the Brain: Neurophotonics Fellowship (Grant No. DGE-1633516).

### Acknowledgment.

The authors acknowledge support from the Boston University Neurophotonics Center. J.G. acknowledges support from Georgia Tech Research Institute.

### Disclosures.

On behalf of all authors, the corresponding authors state that there is no conflict of interest.

### Data Availability.

Data underlying the results presented in this paper are not publicly available at this time but may be obtained from the author upon reasonable request.

### Supplementary.

See Supplement 1 for supporting content

**Supplementary Information Text**

**Impact of Loss and PSF Regularizers on Optimized Results**

We investigate the impact of each loss term on the quality of the learned EDoF, as shown in **Fig. S1**. Each trial was subject to the same reconstruction network, optimization parameters, and same initialization, as discovered by deep genetic initialization. We vary the loss function to contain only MSE (see **Fig. S1A**), all losses (see **Fig. S1B**), and permutations of all losses with our designed regularizers (see **Fig. S1C-F**). The goal of each proposed regularizer is to constrain the physical behavior of the PSF vs the quality of the reconstruction in image space to guide the physical layer into enforcing desired traits. The regularizers proposed in this work include:

1. Peak-to-Peak Variation (PPV) regularization: PPV normalizes the 3D PSF by its maximum value and then draws a line profile at x=y=0 to determine fluctuations in peak intensity along the optical axis. Next, it constrains fluctuations away from the peak value. The objective of this regularizer is to reduce variation along the optical axis by maintaining a comparable amount of optical power on-axis to encourage a consistent EDoF profile. This regularizer takes the form:

$$L_{PPV} = \frac{1}{N_{Z_{EDoF}}} \sum_{z \in Z_{EDoF}} 1 - \frac{PSF(x=0, y=0, z)}{\left(PSF(x=0, y=0)\right)} \quad \text{S1}$$

2. Side Lobe Ratio (SLR) Regularization: SLR constrains the ratio of the per-depth EDoF peak value to the peak side lobe value. The objective of this filter is to reduce the contrast between the engineered EDoF and its side lobes while balancing the achieved extension. We accomplish this regularizer by defining a 2D binary PSF mask ($M$), which blocks the peak (here defined by the full-width-full-maximum of the native unencoded PSF) of the EDoF profile to distinguish peak from side lobe properties. This regularizer takes the form:

$$L_{SLR} = \frac{1}{N_{Z_{EDoF}}} \sum_{z \in Z_{EDoF}} \frac{(PSF(x, y, z) * M)}{PSF(x=0, y=0, z)} \quad \text{S2}$$

3. Peak Enclosed Power (PEP) Regularization: PEP maximizes the amount of power along the optical axis to encouraged increased power in the peak versus side lobes of the optimized profile. The objective of this regularizer is to combine the benefits of PPV and SLR in a single term while relating to an optically motivated quantity of enclosed power. This regularizer uses the same masking strategy as SLR to determine enclosed power of the peak versus the side lobes and takes the form:

$$L_{PEP} = \frac{1}{N_{Z_{EDoF}}} \sum_{z \in Z_{EDoF}} \sum_{x,y} \frac{PSF(x, y, z) * M}{PSF(x, y, z) * (1 - M)} \quad \text{S3}$$

When using solely image space losses (see **Fig. S1A-B**), DeepFilters struggles to properly extend the learned EDoF within the user-defined bounds due to the axial compression on the image plane leading to ambiguity. By adding the direct PSF enforcement of the proposed regularizers, we are able more accurately learn EDoFs within the user defined bounds while imposing the desired properties. In **Fig. S1C**, PPV smooths the profile along the optical axis while encouraging appropriate extension. In **Fig. S1D**, SLR reduces the side lobe presence while encouraging appropriate extension. In **Fig. S1E**, we find that PEP acts as too severe of a regularizer and tends to collabs the EDoF into a learned perfect focus to maximize enclosed power. In **Fig. S1F**, we combine all regularizers to achieve comparable smoothness of **Fig. S1C** with reduced sidelobe presence. The weights for each regularizer were selected by tracking the backpropagated gradient magnitude at the EDoF mask parameters and setting the weight such that each update is approximately equal to the image space loss functions update magnitude.

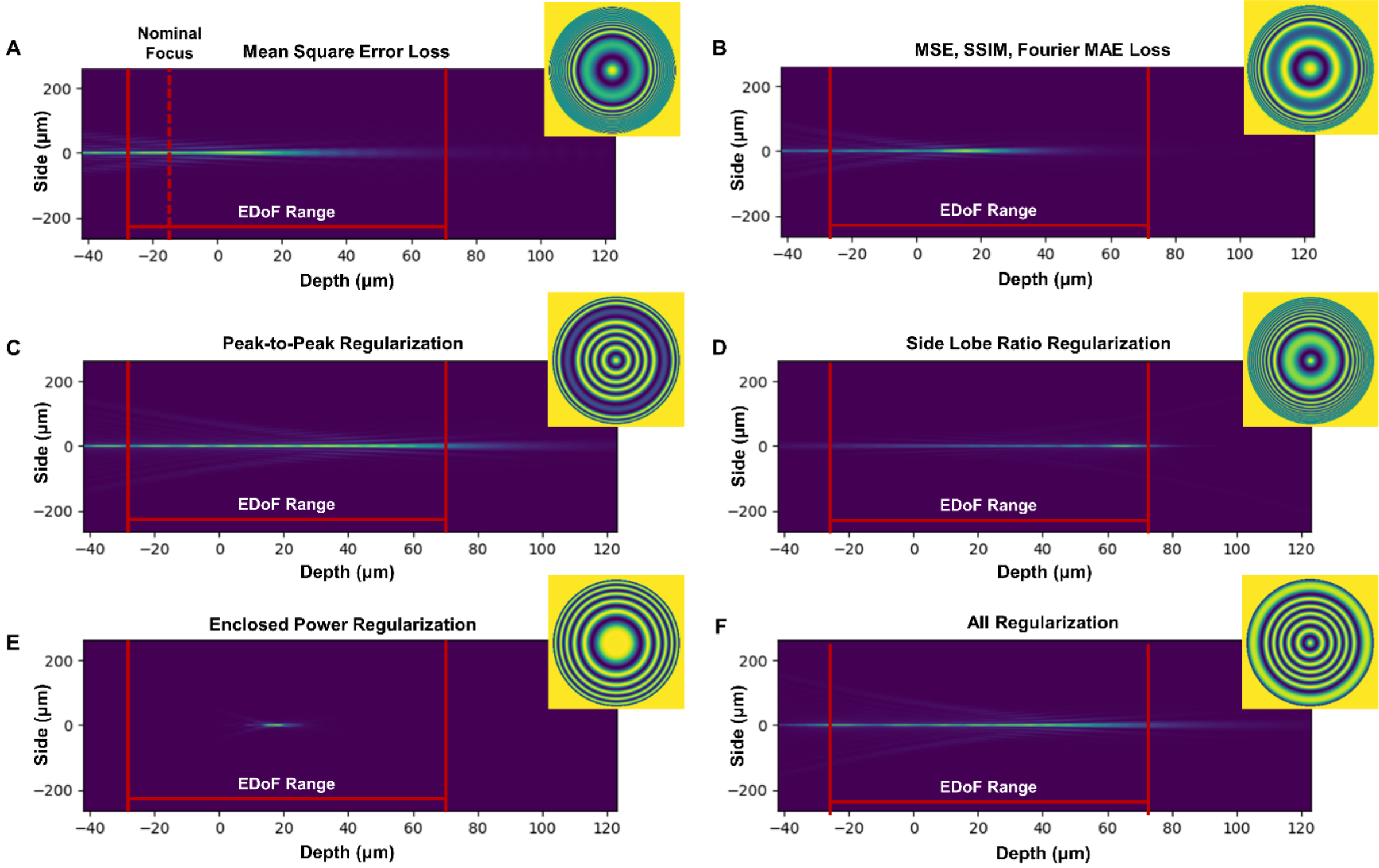

**Fig. S1. Impact of Loss Function on Learned EDoF.** (A) Learned EDoF using image space MSE on a 120 µm desired EDoF as determined by bounds (red solid lines) versus focal plane with native spherical aberration (dashed red line). (B) Learned EDoF using all image space losses. (C) Learned EDoF using all image space losses and PPV. (D) Learned EDoF using all image space losses and SLR. (E) Learned EDoF using all image space losses and PEP. (F) Learned EDoF using all image space losses and all regularizers.

### Impact of the Number of Radial Pupil Basis on Optimization

Next, we investigate the impact of the number of basis terms in the radial polynomial expansion on the achieved EDoF. We optimized for a 120 µm EDoF using a 6-order radial polynominal (see **Fig. S2A**), 8-order (see **Fig. S2B**), and 10-order polynomials (see **Fig. S2C**) and present the results of deep genetic initialization as well as the PSF profile after final optimization. Here, we notice two trends. First, deep genetic initialization achieves lower loss but higher variance in the refined population as the number of parameters scales. The result is intuitive as higher order parameterizations allow for more complex pupil profiles yet introduce an increased number of genes to refine, which increases the complexity of the loss landscape leading to challenges in convergence for evolutionary algorithms [14]. Second, the finalized profile exhibits more variation along the optical axis and reduced EDoF. Similarly, the increased number of parameters leads to more complex optimization and increased variability in the finalized profile. For these reasons, we found that 6-term parameterization serves as an effective tradeoff for EDoF optimization.

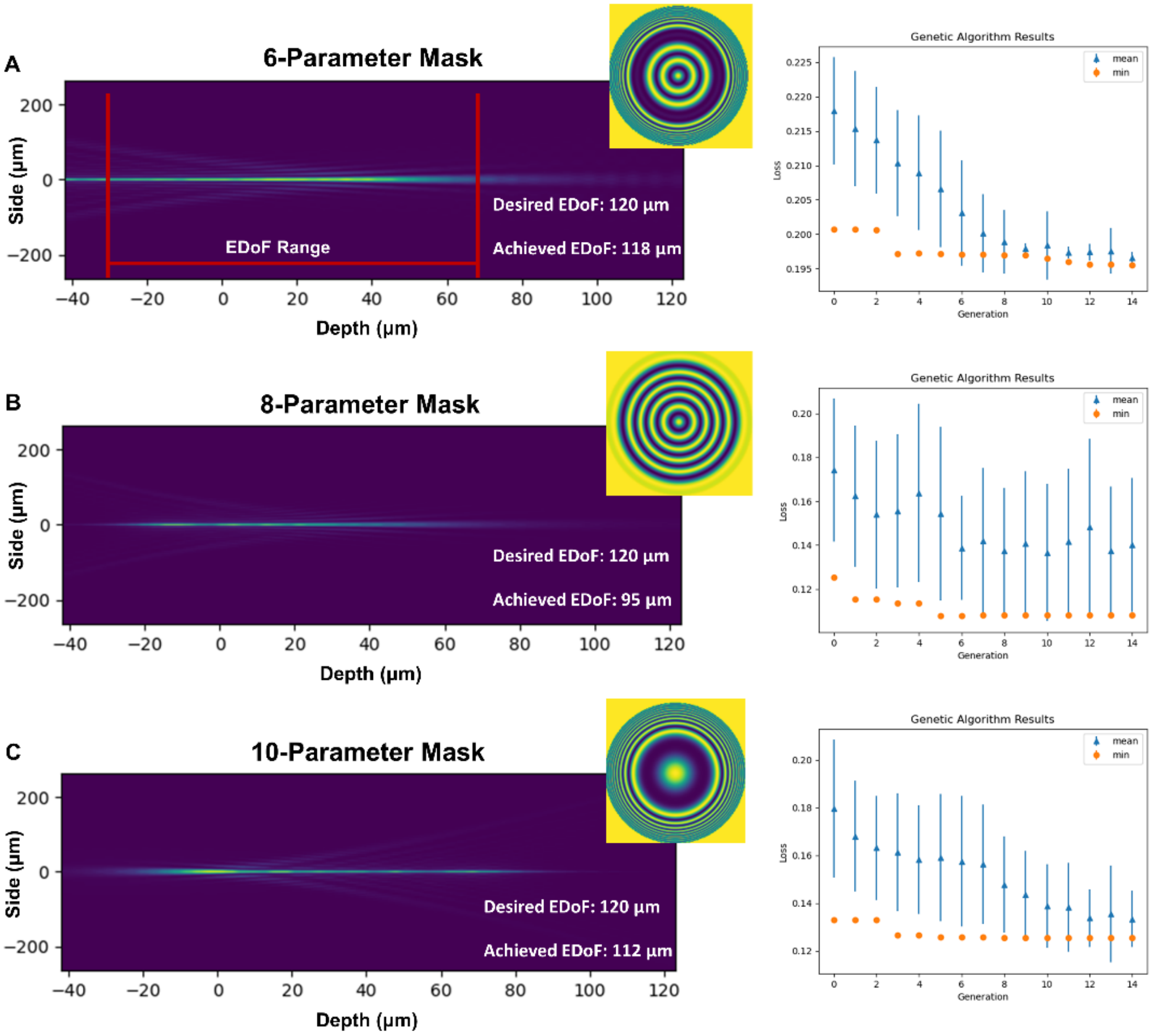

**Fig. S2. Impact of Pupil Parameterization on EDoF Optimization** (A) Deep genetic initialization results (right) and final learned EDoF (right) for a 120 µm EDoF with a 6-term pupil parameterization. (B) Deep genetic initialization results (right) and final learned EDoF (right) for a 120 µm EDoF with a 8-term pupil parameterization. (C) Deep genetic initialization results (right) and final learned EDoF (right) for a 120 µm EDoF with a 10-term pupil parameterization.

### Analysis of Learned Coefficients in Radial Pupil Basis Over Increased Extension

Next, we break down the effects of the learned basis weights over increasing extension lengths. We notice that the basis initially learned favors defocus, spherical aberration, and axicon phase terms, which is intuitive, as multiple projects have selected this combination previously for EDoF optimization [4], [10]. However, as we parameterize a longer extension, we notice two trends. First, the top phase terms receive larger weighting as the optical field requires more tailored shaping to achieve longer EDoFs. Second, axicon is replaced by a 5th-order radial polynomial term. While this term has no native physical association, this result emphasizes how this selection in the basis enables competitive EDoF by merging known EDoF terms with novel associated terms to create an expanded design space to conduct optimization. We show the complete basis breakdown in **Fig. S3**.

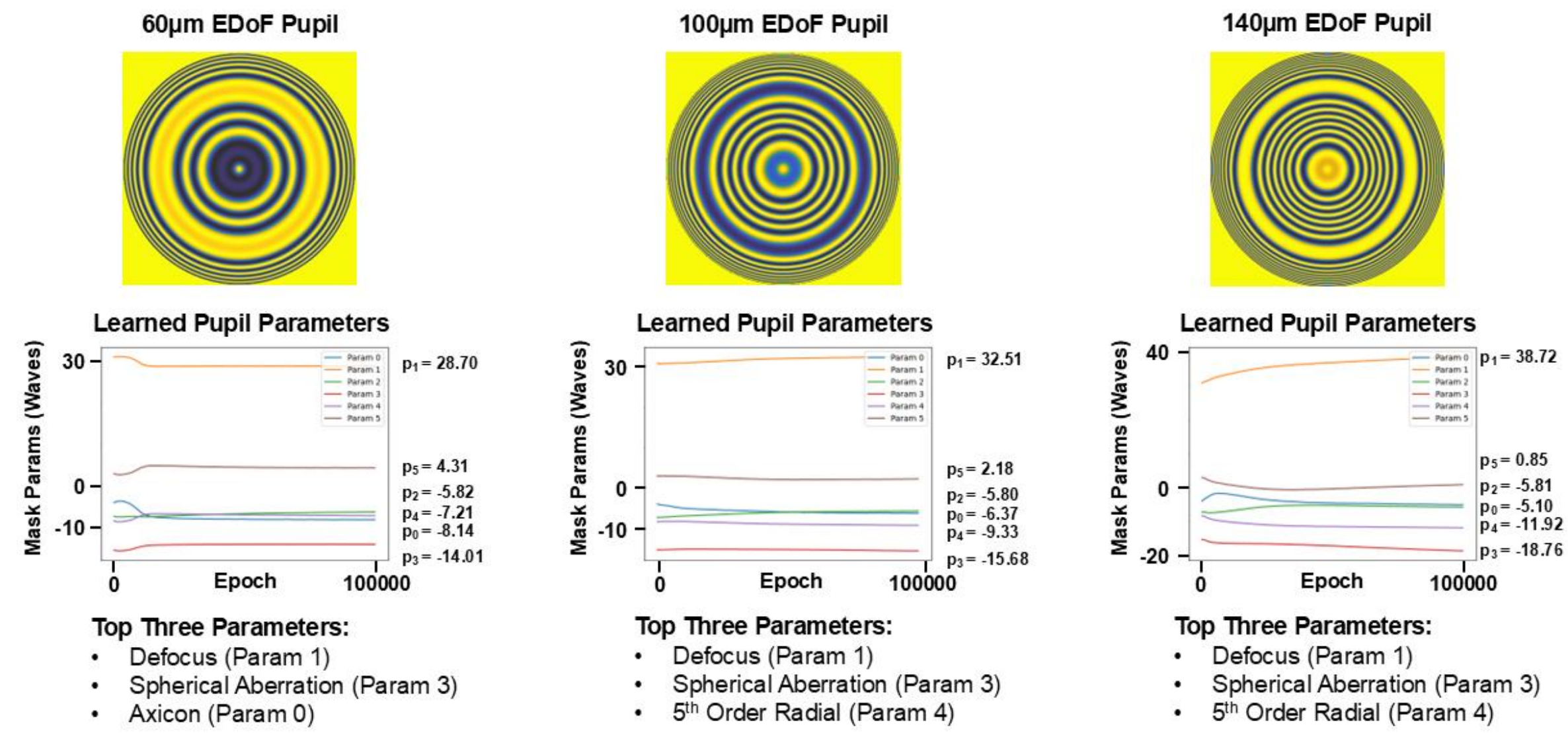

**Fig. S3. Learned Phase Mask Basis Breakdown.** Learned basis terms for simulation for 60 µm, 100 µm, and 140 µm, respectively.

### Impact of Deep Genetic Initialization on Initial Phase Mask Quality

Next, we verify the utility of deep genetic initialization in producing superior initialized pupil candidate populations. We compare the initialization of a population of 30 pupil masks after 10 generations for four different strategies in **Fig. S4**: random, random with genetic evolution, random with deep refinement, and deep genetic initialization. The loss is judged by all image losses, PPV, and SLR regularizers. Each trial attempts to find an initial condition to achieve an initialization for a 100 µm EDoF. This comparison allows us to characterize the contribution of both the genetic- and deep learning-based steps in refining a random population. The random sampling approach randomly creates an entirely new population for each generation, except for the best-performing mask. As expected, this population does not iteratively refine genes and achieves poor convergence and poor quality “best” initial PSF due to the low extension and high sidelobes. By adding a genetic step, the population readily converges to an improved loss but exhibits poor extension. By adding a deep refinement step, the population does not converge but achieves a lower final loss. By combining both a genetic- and deep learning-based step, deep genetic initialization converges and offers the lowest final loss and longest initialized EDoF.

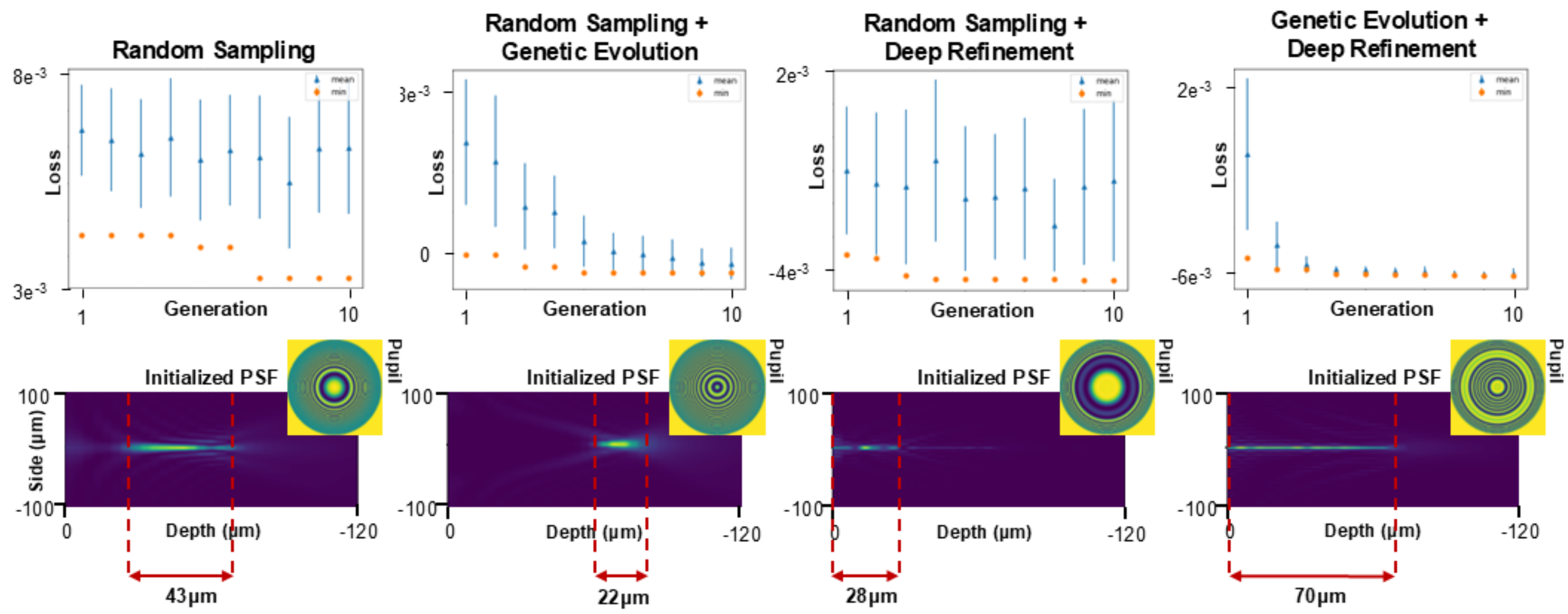

**Fig. S4. Deep Genetic Initialization on Initialized Pupil Populations.** Impact of random sampling alone, random sampling with genetic evolution, random sampling with deep refinement, and deep genetic initialization on finding well-performing initialized conditions from the same initial population.

### Design of a Differentiable and Unrolled Dark Sectioning FilterNet

To support a lightweight algorithm to recover extended fluorescent features, we design a FilterNet based on a modified, differentiable, and unrolled version of dark sectioning [11]. Dark sectioning adapts its methodology from natural image dehazing using a "dark channel prior" to decouple low-frequency object information from background by estimating and removing local patch-wise haze. Here, we innovate on this framework by replacing original operations with differentiable surrogates as well as automating parameters to enable end-to-end learning. We build our method by adapting the original code into a PyTorch nn.Module child class to seamlessly integrate into deep learning workflows. Similar to the observations supported in the original work, we find that unrolling dark sectioning into a two-iteration process to remove background features of varying scales serves as a best practice by allowing for selective parameter learning for global and local background features. We present an overview of our dark sectioning method in **Fig. S5**.

In brief, our dark sectioning FilterNet takes in an image frame from EDoF-Tabletop, $I_{in}$, and a single user-parameter, $d_{block}$, which is the local patch size used to determine background haze. In general, this parameter is set to the approximate feature size for the underlying scene. Next, the input image is filtered with a filter bank of three Gaussian filters: a high-pass filter (HI), a low-pass filter (LO), and an extreme low-pass filter (ELO). In this formulation, HI is designed to contain weak high-frequency object information free from background, LO is designed to contain a mixture of object and background information, and ELO is designed to contain purely background information. Since LO contains both object and background features, we may treat LO as the fusion of a low-pass background map, BG, and low-pass object map, $LO_{IF}$, through a transmission function, T:

$$LO(x,y) = LO_{IF}(x,y) \cdot T(x,y) + (1 - T(x,y)) \cdot BG(x,y) \qquad \text{S4}$$

where BG is estimated through ELO. LO is set according to the optical transfer function (OTF) cutoff, and HI is the reciprocal filter. We set the standard deviation, $\sigma_{LO}$, to be 0.5x the OTF cutoff in the first iteration in accordance with the original work, but find that constraining $\sigma_{LO}$ to a lower value (e.g. 0.3x the OTF cutoff) allows for improved estimation of remaining background haze during the second iteration. ELO is either set as a learned parameter in the end-to-end experiments or set below $\sigma_{LO}$ in hand-tuned experiments. Next, we use LO to determine the maximum greyscale values cutoff associated with the out-of-focus haze, $\Delta_{OOF}$, and in-focus object features, $\Delta_{IF}$. To determine these values, we first create a thresholded version of LO using a threshold value, $\tau_{LO}$, to act as an initial background separation point. In the original work, this value is set by inspection, whereas our version uses the mean value of LO. Background estimation is sensitive to this threshold, and we find that the use of the mean reduces complexity but leaves haze surrounding dense object regions, partially contributing to the necessity of a second iteration of dark sectioning to fully remove background. Next, both the thresholded LO image, LO$\tau_{LO}$, and LO are sent through dark channel filter. Dark channel replaces a local pixel with the minimum value across all color channels within a patch set by the block size to approximate the background. In practice, we replace this operation with a channel-sensitive erosion filter as a differentiable alternative. Next, we average the top 1% of values in each image histogram to determine the maximum value while reducing the impact of outliers. We use these values to renormalize ELO into the initial background estimate, BG, such that the approximated background spans from the out-of-focus background value to the in-focus background value by:

$$BG(x,y) = \delta \frac{ELO(x,y) - (ELO(x,y))}{(ELO(x,y)) - (ELO(x,y))} * (\Delta_{IF} - \Delta_{OOF}) + \Delta_{OOF} \qquad \text{S5}$$

Where $\delta$ is a scaling term used to over- or under-suppress the background in the next step, which is set to 3 for the first iteration based on the original work and set to 0.95 for the second iteration. Next, by taking **Eq. S4** and dividing BG on each side and performing a dark channel operation, we may estimate:

$$dark(\frac{LO(x,y)}{BG(x,y)}) = dark(\frac{LO_{IF}(x,y)}{BG(x,y)} \cdot T(x,y) + \left(1 - T(x,y)\right)) \qquad \text{S6}$$

Next, we impose that the dark operation erodes the support of each side of **Eq. S6** by the feature size. Under this operation, we may assume that $O_{LO} \sim 0$ as all feature support is eroded, leading to:

$$dark\left(\frac{LO(x,y)}{BG(x,y)}\right) \sim dark(\left(1 - T(x,y)\right)) \qquad \text{S7}$$

Next, we define $dark\left(1 - T(x,y)\right) \sim 1 - C(x,y)$, which is a local low-resolution background contrast map. To restore the original transmission function, we perform a guided filter operation to restore smooth details of C based on the structure of LO by:

$$T(x,y) = \frac{cov\left(LO(x,y),\ C(x,y)\right)}{var\left(LO(x,y)\right) + \varepsilon} \cdot \left(LO(x,y) - \underline{LO}(x,y)\right) + \underline{C}(x,y) \qquad \text{S8}$$

Where $cov$ is the covariance operation, $var$ is the variance, $\varepsilon$ is a small scalar to prevent divide by zero, and $\underline{\cdot}$ is the mean. Intuitively, the first term determines the slope of a least-squares regression, the second term applies it to the normalized variation around the mean, and the third term adjusts the regression line around the mean of the contrast map. In our implementation, we use a patch-wise averaging pooling operation to rapidly calculate the required quantities while supporting differentiability. Now that T is determined, we may recover the true low-pass object information by inverting **Eq. S4** to produce:

$$LO_{IF}(x,y) = \frac{LO(x,y) - (1-\omega)\big(1-T(x,y)\big)\cdot BG(x,y)}{max\ (T(x,y),\ \beta)} \cdot (I_{in}) \tag{S9}$$

Where $\beta$ is a hard threshold set to 0.1 to prevent division by zero and $(I_{in})$ rescales the image to the input image depth. $\omega$ is a blending term to retain a fraction of the predicted background to prevent oversuppression of object information. The final image, $I_{out}$, is synthesized by merging this estimate with HI:

$$I_{out}(x,y) = LO_{IF}(x,y) + HI(x,y) \tag{S10}$$

**A**

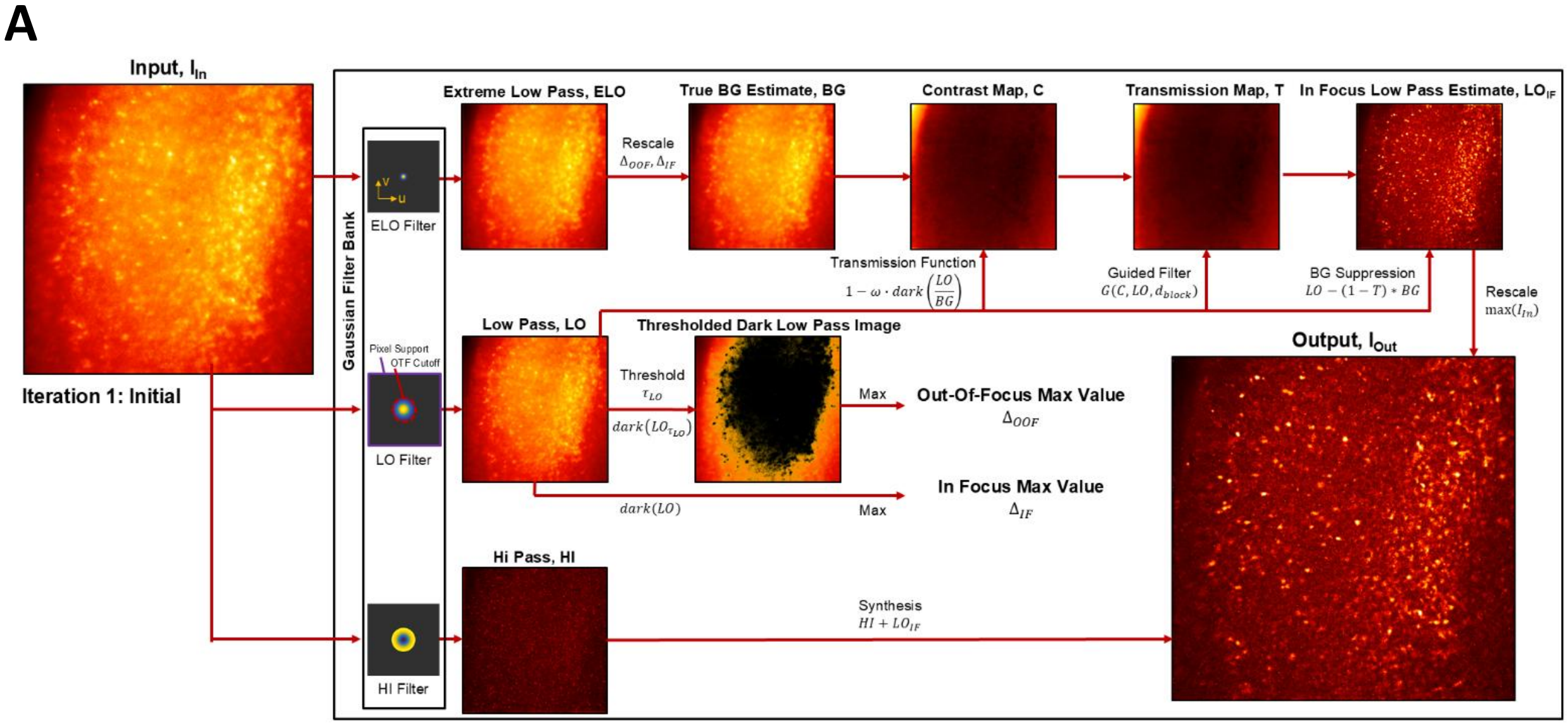

**B**

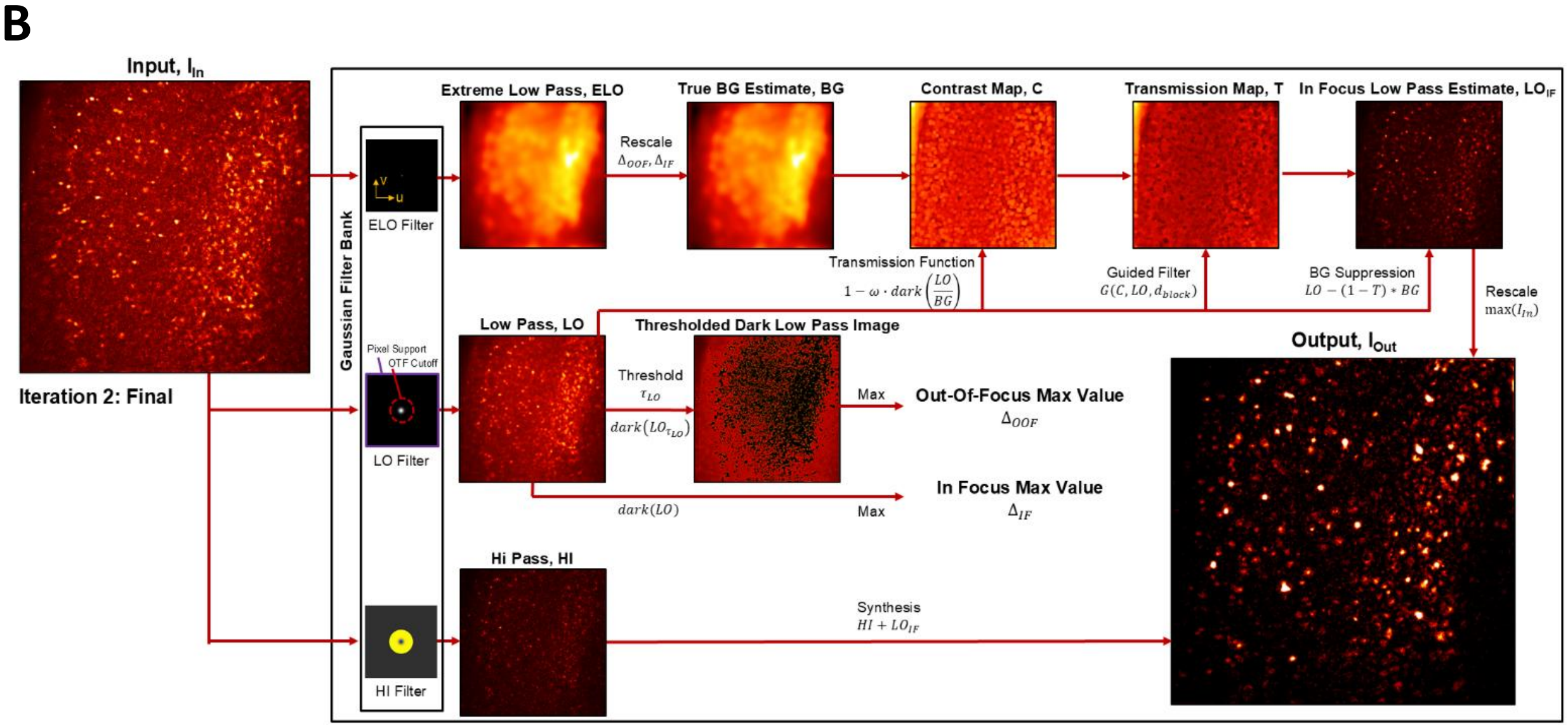

**Fig. S5. Unrolled Dark Sectioning Overview. Overview of a two-iteration ((A) Course, (B) Fine) differentiable dark sectioning pipeline used as a FilterNet.**

## One-Time Calibration of Objective Spherical Aberration

While traditionally seen as undesirable due to lower image contrast, understanding and controlling known aberrations serves as an effective method for engineering effective EDoFs [4], [10]. To merge traditional and novel aberration terms, the method explored in this work utilizes a novel radial basis on the pupil plane to map to known EDoF solutions (e.g. axicon, defocus, cubic phase, spherical aberration) with options to extend to unexplored radial orders (such as $r^5$). To appropriately predict the EDoF produced by EDoF-Tabletop, we must perform a one-time calibration to characterize the native amount of elongation present within the system before optimization. Here, we assume that the native elongation is dominated by spherical aberration due to the use of high NA spherical lenses and may be appropriately absorbed as a calibrated parameter in our prescribed basis once determined. To perform the calibration, we prepare a sparse 1 µm green polystyrene fluorescence bead (Thermo Fisher F13081) sample and isolate a single bead on the optical axis of EDoF-Tabletop. Next, we sweep the bead axially between $\pm$50 µm of the focal plane to capture the axial profile, as shown in **Fig. S6A-B**. Next, we synthetically generate spherically elongated PSFs using the parameters of the system to match the degree of elongation, as shown in **Fig. S6C-D** to determine the appropriate amount of aberration. We use the inflection of spherical aberration around the focal point to determine the polarity of the spherical aberration terms.

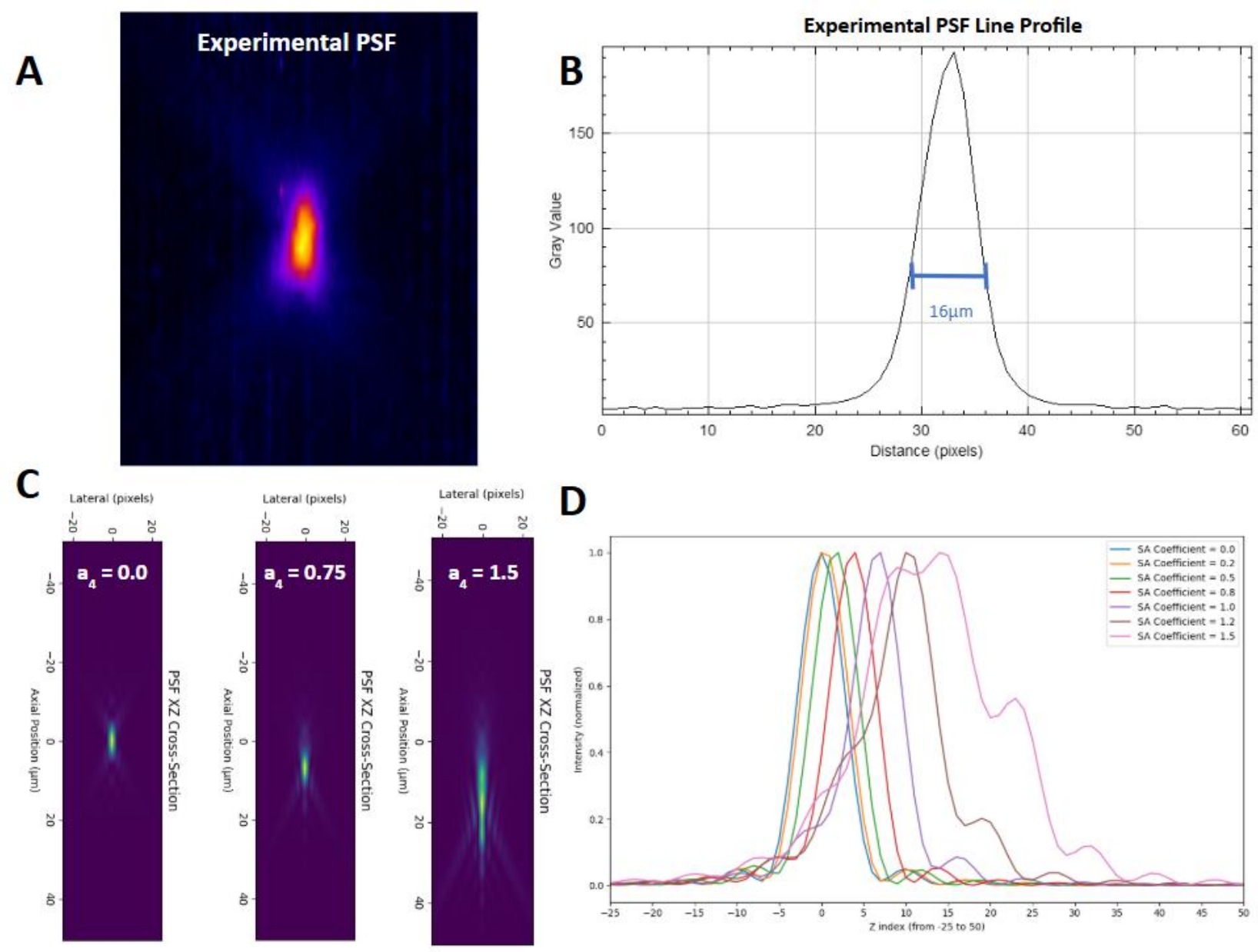

**Fig. S6. Spherical Aberration Calibration of EDoF-Tabletop.** (A) Measured PSF of EDoF-Tabletop. (B) Line profile of PSF along the optical axis to measure elongation. (C) Simulated PSFs to match elongation. (D) Elongated XZ PSF cross section versus spherical aberration.

## Performance of EDoF-Tabletop on Whole Fixed Brain Vasculature

We demonstrate the recovery of extended vascular networks from a whole intact fixed mouse brain. This experiment builds on our prior results imaging tissue fibers and neural slices with the intention to validate DeepFilters' capability to recover non-cellular structures deep from complex biological tissue. We showcase the depth-encoded MIP of the interrogated vasculature network in **Fig. S7A** as well as the progressive FilterNet recovery for PSFs with nominal length from 200µm to 400µmWe present a reference image of the brain sample in **Fig. S7D**. Overall, we recover vascular features down to 400 µm depth with significantly improved contrast. This demonstrates that DeepFilters can process complex biological data across diverse morphologies without retraining or architectural modification.

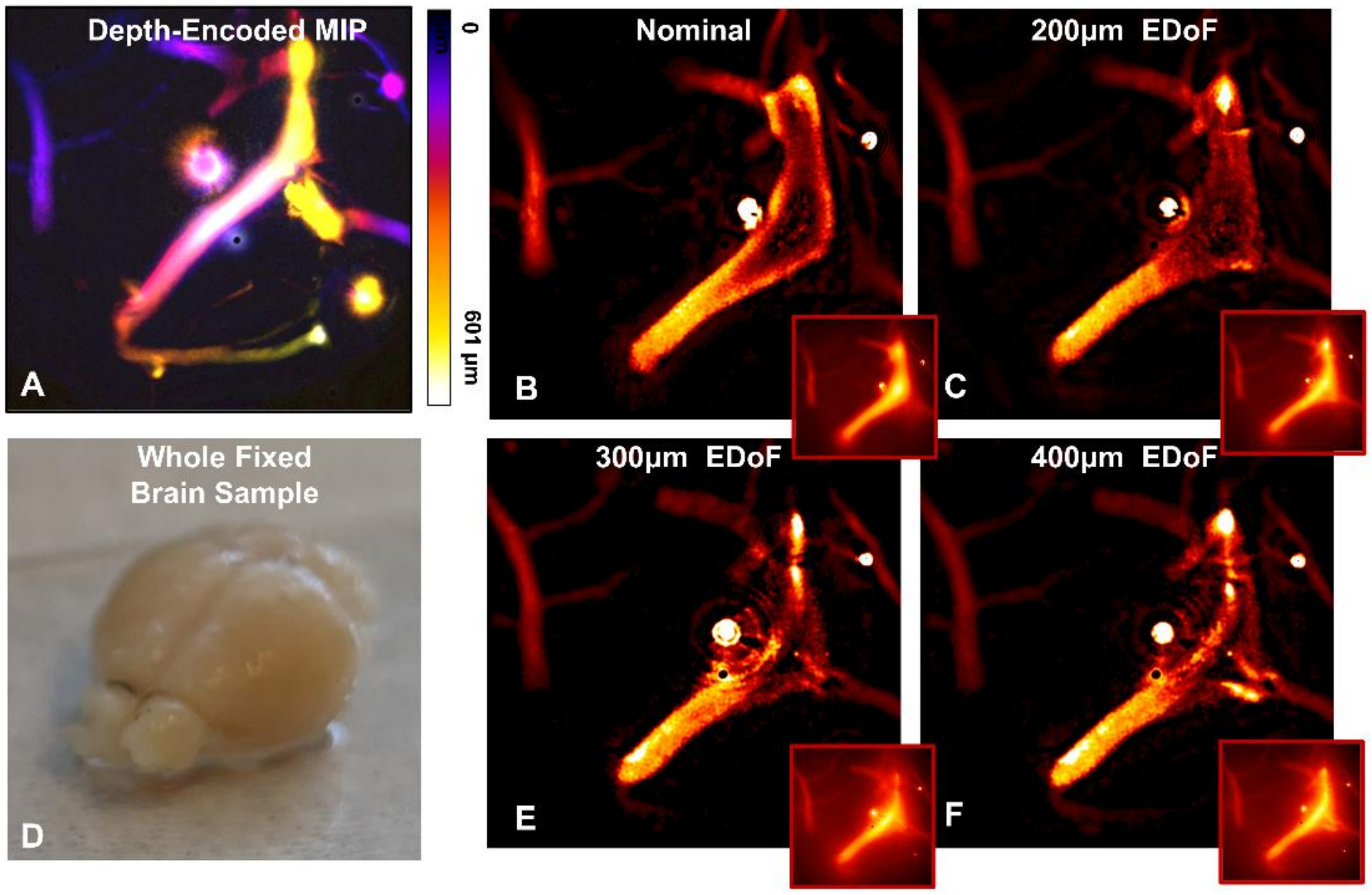

**Fig. S7. Extended Vascular Recovery in a Whole Fixed Brain.** (A) Depth encoded MIP of vasculature in a whole fixed brain sample. (B) recovery of vasculature after FilterNet processing versus raw data (inlet) for nominal PSF. (C) Repeat for 200 µm EDoF. (D) Image of the whole fixed mouse brain sample used in the experiment. (E) Repeat analysis for 300 µm EDoF. (F) Repeat for 400m µm EDoF.

## Comparison of Simulation of EDoFs under Proxy Scattering Models

Here, we analyze the robustness of optimized EDoFs to the effects of scattering by comparing our proxy scattering kernel to prior work that developed a split-step Fourier optics-driven approach simulating brain tissue with g=0.9 and ls=100 µm[13]. First, we simulate a variety of different beams under scattering considerations and measure the reduction in DoF due to scattering, as shown in **Fig. S8A-C**. We simulate an unmodulated and unaberrated PSF in **Fig. S8A**, an axicon encoded PSF in **Fig. S8B**, and an optimized PSF mask in **Fig. S8C**. We notice that both the unencoded PSF and axicon PSF exhibit a comparable relative reduction in DoF between free space (top row) and scattering (bottom row) profiles of 17.8% and 24.7%, respectively. This result indicates that while traditional methods may achieve additional elongation in scattering media, even assumed "self-healing" beams, such as axicon, struggle in strongly scattering environments to achieve their full DoF. In contrast, our optimized result achieves over 2x the elongation of the simulated axicon PSF yet only experiences a relative 7.8% reduction (see **Fig. S8C, Blue Line**) in elongation between free space (top) and scattering (bottom) profiles. This result confirms that our scattering-aware optimization approach is successful in optimizing profiles that are both sufficiently elongated and resilient for utility in highly scattering applications, such as biological imaging. Next, we compare this high-fidelity split-step model to the proxy scattering model used in this work. We find that the free space simulations match closely despite differences in resolution (see **Fig. S8D**). However, we find that our scattering simulation overapproximates the effects of scattering, as shown in **Fig. S8E**. We believe this discrepancy is due to calibrating the scattering kernel on an unencoded PSF, which is more subject to scattering-based degradation. In future works, we will investigate other methods to capture split-step effects in the end-to-end learning pipeline.

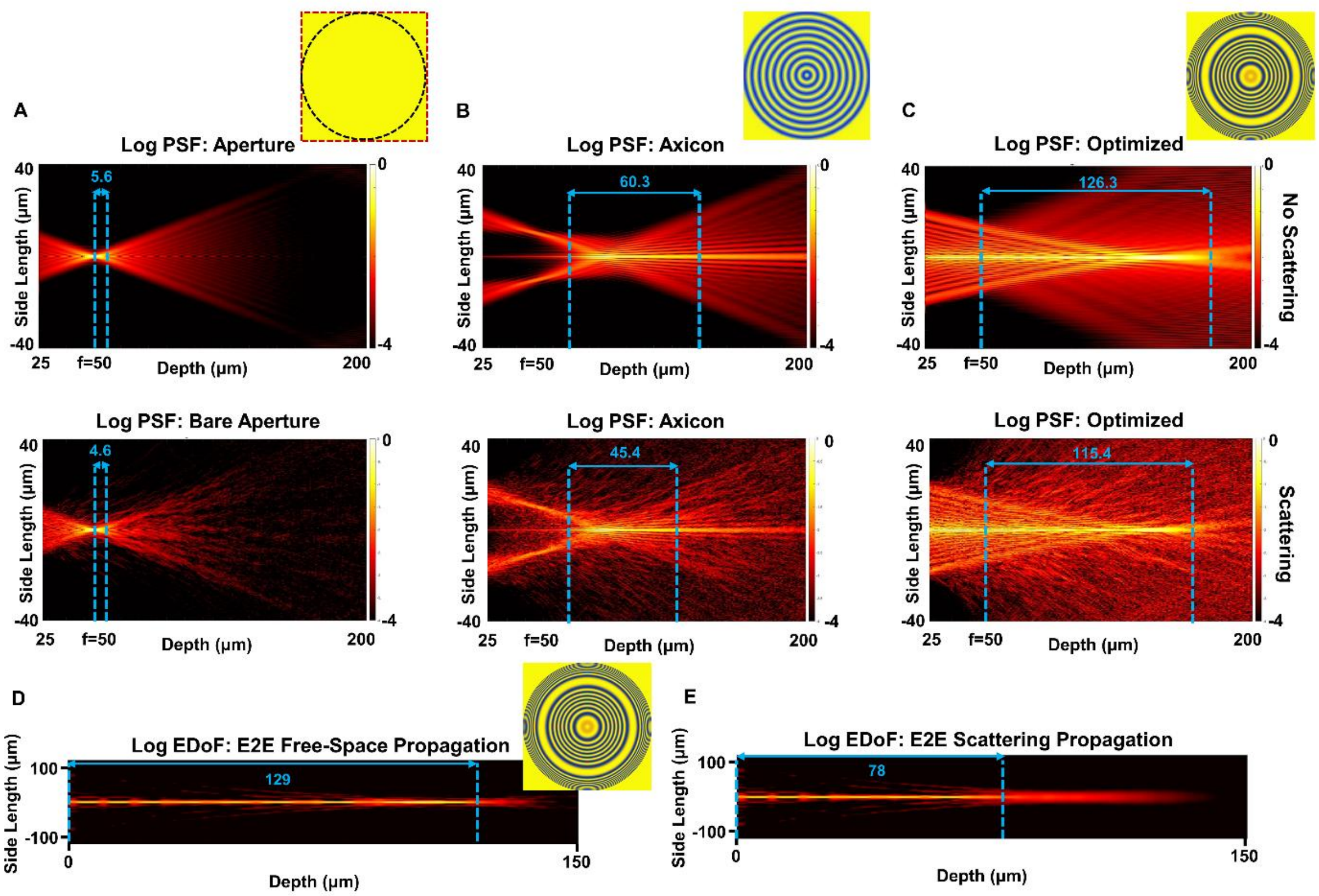

**Fig. S8. EDoF Resilience and Comparison to Split-Step Simulation.** (A) Split-step Monte Carlo simulation of an unencoded PSF without (top) and with (bottom) scattering. (B) Split-step simulation of an axicon PSF without (top) and with (bottom) scattering. (C) Split-step simulation of an optimized PSF without (top) and with (bottom) scattering. (D) Free space PSF of the DeepFilters forward model of the optimized PSF. (E) Scattered PSF of the DeepFilters forward model of the optimized PSF.

## End-to-End Training with FilterNets

Here, we demonstrate the capability to train FilterNets in an end-to-end framework to merge the training and deployment steps. We set up a FilterNet consisting of an adaptive media filter layer and unrolled dark sectioning layer to serve as the reconstruction layer in the deep optics pipeline described in the main text. To select which parameters to learn, we recognize that the parameters underlying FilterNets contain more physical intuition than conventional neural networks and may be generally set manually by inspecting data features or by intuition. For the proposed FilterNet, we may set most of the parameters by our object size (e.g., neurons, fibers, vessels) and by consulting the original work [11].

| FilterNet Parameter | Initial | Learned |
|---|---|---|
| Adaptive Median Filter Min Kernel | 3 pixels | N/A |
| Adaptive Median Filter Max Kernel | 7 pixels | N/A |
| Sectioning Low Pass Waist ($\sigma_{LO}$) | [0.3, 0.3] | [5.98, 5.45] |
| Sectioning Extreme Low Pass Waist ($\sigma_{ELO}$) | [0.05 0.1] | N/A |
| Sectioning Block Size | [10, 10] pixels | N/A |
| Sectioning Background Rejection Factor ($\omega$) | [0.95, 0.95] | [0.775, 0.541] |
| Sectioning Background Estimate Scale ($\delta$) | [3, 0.95] | [1.98, 0.66] |

**Table 1. FilterNet Tunable Parameters.** Parameters used to process data using the described FilterNet. Initial values are handset. Bracket values indicate values for each of the two unrolled dark sectioning iterations. Learned values are set as learnable end-to-end.

However, we recognize that the extreme low-pass background estimate scaling term, $\delta$, is difficult to tune in our modified framework due to the automation of the assumed background greyscale value, leading to arbitrariness for a given bit-depth. As such, we decided to set this parameter as learnable during training. For training, we follow the same procedure described in the main text but tune the learning rate such that the gradient update remains comparable in value after backpropagation through the FilterNet and physical simulator. In addition, due to the lower number of network parameters to train, we find that we do not require as many epochs to converge, and we can decrease the training length from 100,000 to 15,000 accordingly compared to using a UNET architecture for the reconstruction. We present the results in **Fig. S9.** As shown in **Fig. S9A-B**, the learned FilterNet parameters (displayed in **Table 1** with the remaining FilterNet parameters) outperforms the hand-tuned FilterNet in terms of background suppression, here demonstrated on the 200 µm fixed mouse brain slice. Alongside superior visual performance, we note that the end-to-end training allows for a reduction in the image-space loss functions (see **Fig. S9C**) and PSF regularizers (see **Fig. S9D**), indicating for balanced convergence in the network and physical parameters. When inspecting the learned PSF, we noticed comparable elongation than achieved through the U-Net trainer, up to a 400 µm EDoF, with reasonable confinement within the user-defined depth region, as shown in **Fig. S9E-G**.

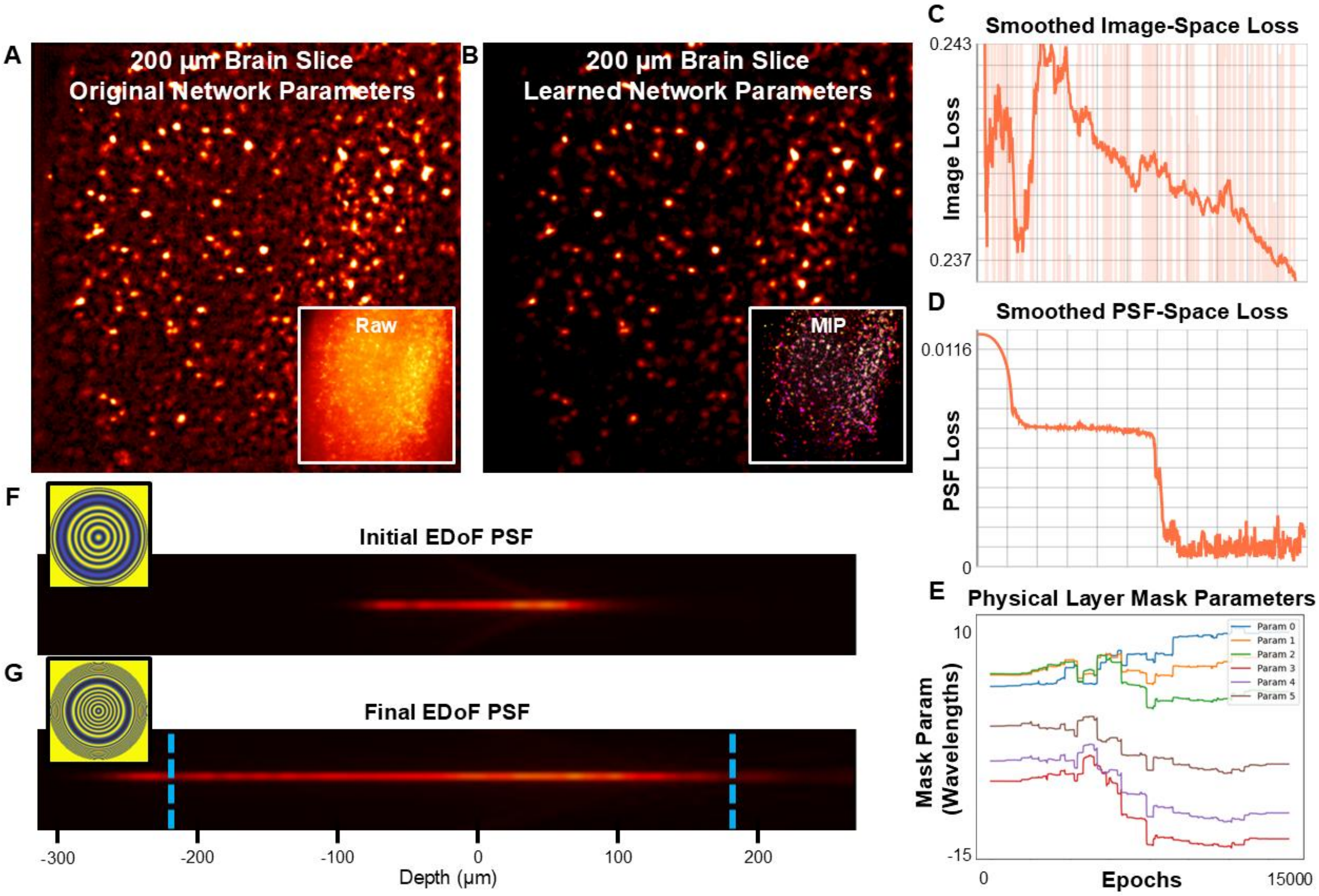

**Fig. S9. End-to-End FilterNet Training.** (**A**) Initial 200 µm fixed mouse brain reconstruction with hand-tuned and (**B**) learned parameters. (**C**) Image-space loss and (**D**) PSF-space regularizer values during training. (**E**) Learned pupil filter parameters to create elongation from (**F**) initial EDoF to (**G**) user-defined EDoF.

### Matching the Forward Model to Physical Simulations

To encourage accurate simulation, we utilize a Fourier optical-based forward model to impose the wave effects of a modulated pupil on a synthetic example during training. We use the modeling and proxy scattering kernel described in the main text and detailed in **Fig. S10**. To ensure accuracy, we characterize the power spectral density (PSD) of our synthetic neural examples against real fixed brain samples. To control the spatial frequency composition, we scale our synthetic examples such that the neurons exhibit comparable size to real mouse brains neurons (~4-20 µm) and use a Poisson-Gaussian noise model calibrated based on our camera and value noise-inspired background to inject realistic effects. We use prior work to appropriately scale each term for our forward model [12]. We balance the PSD across multiple samples of differing thickness to ensure accuracy of the forward model in replicating neural scenes.

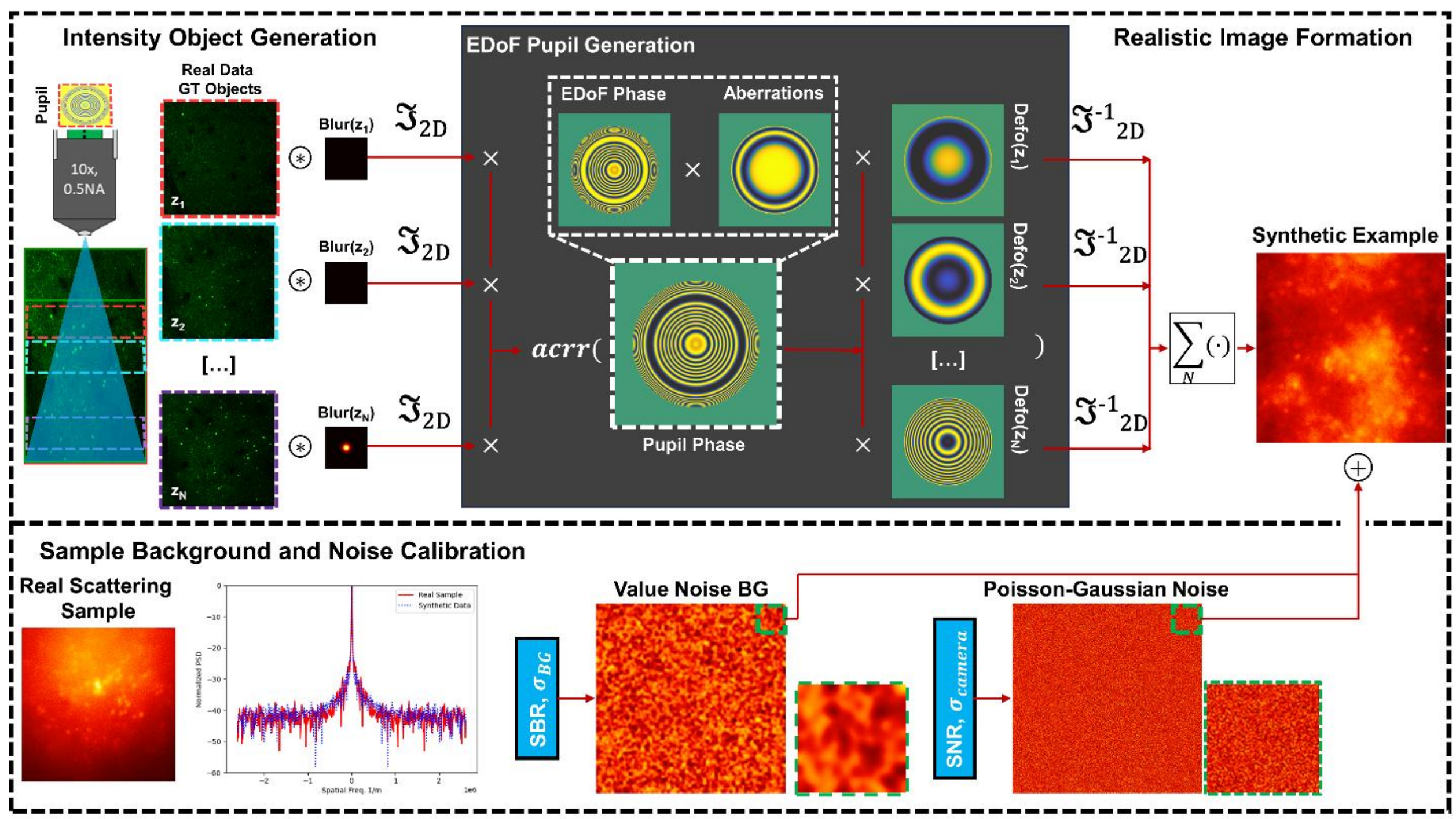

**Fig. S10. Forward Model Calibration**. We use a value-noise and shot-noise modified Fourier optics simulation to match the PSD of synthetic neural images to real experiments to minimize mismatch in optimized results.

**Comparison of Results Under Different Reconstructions**

To further verify the utility of our end-to-end trained FilterNets, we perform a meta-analysis of the performance of our trained dark sectioning method against several other processing techniques. We compare against our trainer U-Net as well as other filter banks motivated for biological imaging, such as M.O.R.E. filtering [15] (adaptive Median filtering, morphological Opening, Rolling ball background subtraction, Enhancement), and wavelet-based background and noise subtraction (WBNS) [16] algorithms. We compare the results from each method to the collected MIP reference for each experiment and calculate MSE, peak-SNR, and SSIM to judge their fidelity. We present the results in **Fig. S11**.

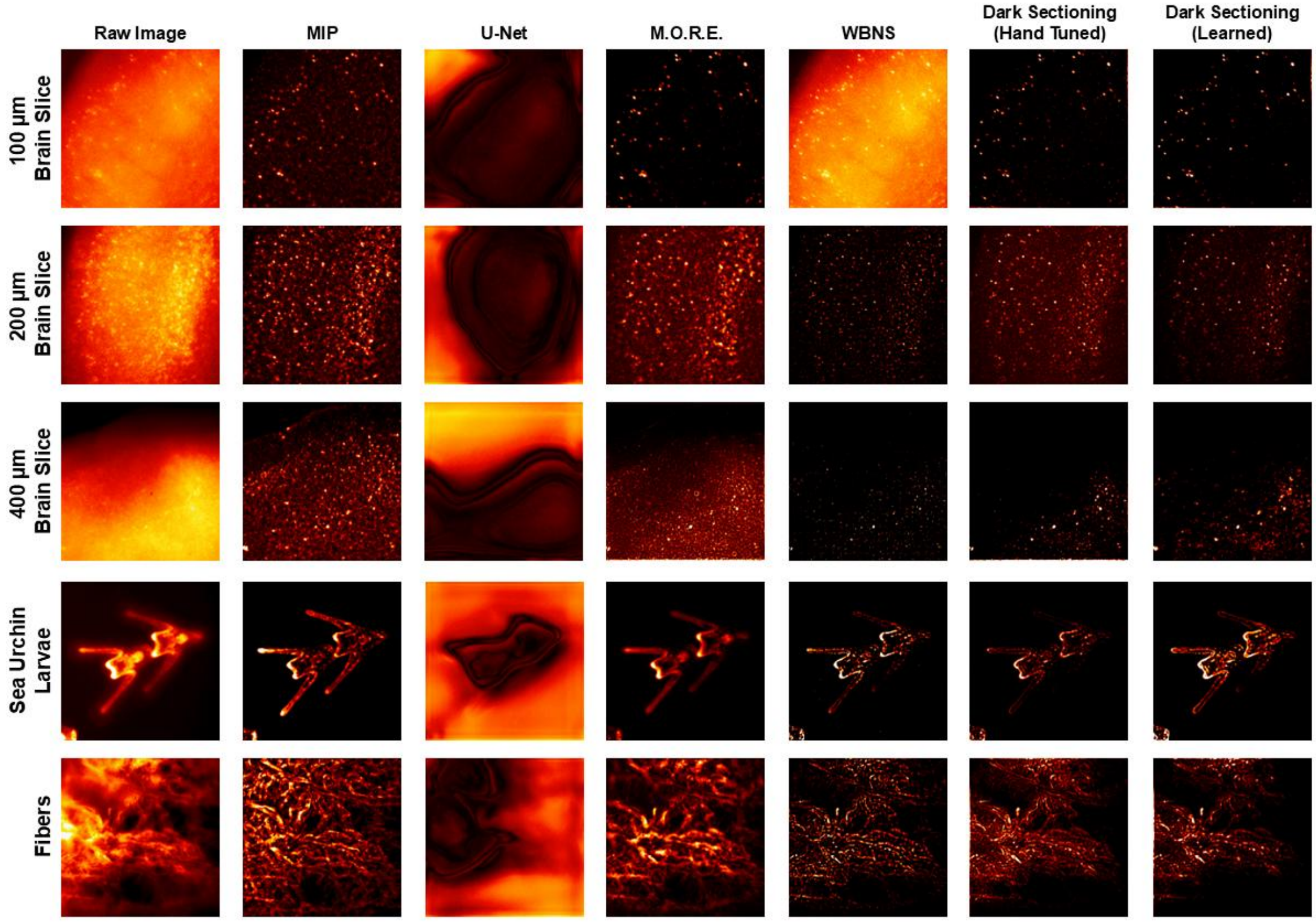

**Fig. S11. Comparison of Reconstruction on Experimental Data.** Comparison of U-Net, M.O.R.E. filtering, WBNS, and the hand-tuned and trained dark sectioning methods across experimental data.

Next, we aggregate the performance metrics across all experimental trials and present their average in **Table 2** for each reconstruction method. Notably, the trainer U-Net is outperformed by all wavelet-based methods due to the susceptibility to data drift. In terms of filter-based methods, dark sectioning outperforms the other methods with the end-to-end trained variant on dark sectioning producing the best results. This study supports the necessity of FilterNets for generalized deployment and the ability of end-to-end trained FilterNets to achieve leading performance, despite being similarly trained on synthetic data subjective to drift.

| | U-Net | M.O.R.E. | Wavelet | DS | DS_L |
|---|---|---|---|---|---|
| MSE | 0.14783 | 0.01321 | 0.00939 | 0.00776 | 0.00736 |
| PSNR | 8.81 dB | 19.86 dB | 21.82 dB | 22.94 dB | 23.05 dB |
| SSIM | 0.1952 | 0.6125 | 0.6146 | 0.6579 | 0.6707 |

**Table 2. Quantitative results of reconstruction methods.**

### Convergence Analysis for Different EDOF Ranges

We performed a convergence analysis by repeating the E2E optimization for 10 different seeds, keeping all hyperparameters the same across the runs. This is to test the robustness of the pipeline to the inherent stochasticity of gradient optimization and data-driven approach. To quantify performance, we tracked the learned phase mask coefficients and dark sectioning parameters (namely the contrast scaling and low-pass cutoff terms), the learned EDOF extension range, compound image and PSF losses (as defined in the main text), and various image loss metrics (MSE, PSNR, PCC, MS-SSIM). We repeat this for 5 different EDOF ranges (200, 400, 600, 800, 1000μm). The 1/e dropoff can be further optimized by tweaking the $\alpha$ parameter for balancing the loss for the waist of the EDoF profile with respect to the EDoF cutoff. The 0.1 falloff is shown for empirical demonstration of the target extension. The results are summarized in the tables below. The mean and standard deviation of the learned parameters are used in the following section to probe the sensitivity of the optimized result to the variability in the learned values.

The statistical spread in the learned values has negligible effects on the imaging and reconstruction performance. Note that the 1/e extension range is merely a heuristic for quantifying performance ad hoc, and there is no explicit definition of this range within the optimization. The "0.1 fall-off range" is included to illustrate the extension of the PSF up to the desired range. The 600μm and 800μm trials seem to underextend and overextend their targets, respectively. This can be addressed by tuning the hyperparameter that scales the PSF loss in the EDOF and BG regions, effectively softening the loss behavior at the boundary to encourage some extension below and beyond the hard cutoff value.

| | 200μm | | 400μm | | 600μm | | 800μm | | 1000μm | |
|---|---|---|---|---|---|---|---|---|---|---|
| Parameter | Mean | Std | Mean | Std | Mean | Std | Mean | Std | Mean | Std |
| LPc | [6.166, 5.939] | [3.814E-6, 4.768E-7] | [6.166, 5.939] | [3.814E-6, 4.768E-7] | [6.166, 5.939] | [3.814E-6, 4.768E-7] | [6.166, 5.939] | [3.814E-6, 4.768E-7] | [6.166, 5.939] | [3.814E-6, 4.768E-7] |
| elo | [1.996, 0.756] | [1.19e-6, 0] | [1.996, 0.756] | [1.192E-6, 0] | [1.996, 0.756] | [1.192E-6, 0] | [1.996, 0.756] | [1.192E-6, 0] | [1.996, 0.756] | [1.192E-6, 0] |
| a0 | 3.659 | 6.559E-5 | 3.2214 | 1.04E-03 | -17.5716 | 0.018384 | 17.1993 | 3.73E-15 | -37.4193 | 0.225371 |
| a1 | 5.816 | 5.545E-5 | 23.751 | 8.39E-04 | -4.2861 | 0.005648 | -6.5249 | 9.32E-16 | 5.794075 | 0.194506 |
| a2 | 5.846 | 7.038E-5 | -3.339 | 4.97E-04 | 6.9297 | 0.000256 | -14.553 | 1.86E-15 | 7.137801 | 0.302211 |
| a3 | -6.635 | 1.893E-5 | -20.51 | 1.40E-04 | 1.7777 | 0.002454 | -17.478 | 3.73E-15 | 13.9521 | 0.239968 |
| a4 | -4.78 | 2.696E-5 | -10.56 | 1.40E-04 | 2.5419 | 0.004045 | -7.7095 | 1.86E-15 | 14.2093 | 0.074955 |
| a5 | -1.28 | 4.036E-5 | 4.6490 | 3.59E-04 | 18.386 | 0.005174 | 17.1993 | 3.73E-15 | 15.42907 | 0.096318 |

**Table 3. Learned E2E parameters across different target EDOF runs**

| **Extension Target [μm]** | Extension (1/e falloff) [μm] | Extension (0.1 falloff) [μm] | loss_img | loss_psf | MSE | PSNR | C_PCC | MS-SSIM |
|---|---|---|---|---|---|---|---|---|
| **200** | 163 | 202 | 0.1857 | 0.00553 | 0.009851 | 20.49152 | 0.421336 | 0.632657 |
| **400** | 332 | 393 | 0.1905 | 0.007904 | 0.009851 | 20.49152 | 0.421336 | 0.632657 |
| **600** | 411 | 552 | 0.1793 | 0.008472 | 0.009851 | 20.49152 | 0.421336 | 0.632657 |

| **800** | 681 | 926 | 0.1728 | 0.005347 | 0.009851 | 20.49152 | 0.421336 | 0.632657 |
|---|---|---|---|---|---|---|---|---|
| **1000** | 711.5 | 998 | 0.1825 | 0.007471 | 0.009851 | 20.49152 | 0.421336 | 0.632657 |

**Table 4. Performance metrics across different target EDOF runs**

For a sensitivity analysis, we set the range for the sampled values given the mean and standard variation from the convergence analysis runs and observe the effects of this variability on the loss. For instance, if the convergence analysis gave a range of 3.659±6.559E-5 for the learned $a_0$ mask coefficient value, we randomly select values within one standard deviation of the mean. We sample $a_0$-$a_5$ coefficients simultaneously for 20 trials as a simple Monte Carlo-esque analysis on the PSF extension result. The metrics are compared to the base results (computed using the mean values). This can be repeated for any specific parameter of interest. Even for the most difficult EDOF extension case (1mm), the variation in PSF extension is negligible. This proves that the pipeline is reproducible under the correct conditions, wIth the caveat that the extension into the BG region can be finely tuned by relaxing the loss scaling hyperparameters.

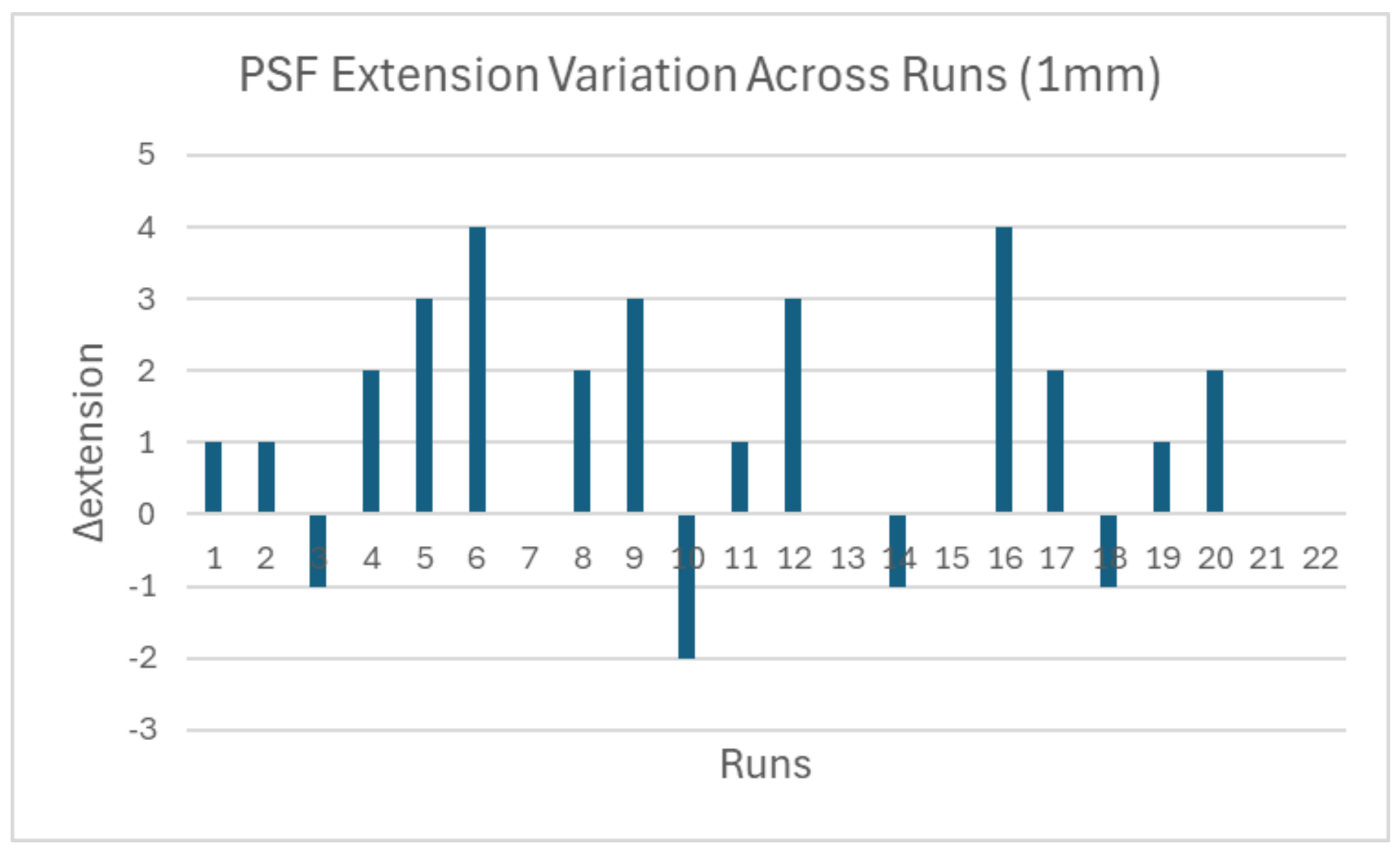

**Fig. S12. Example sensitivity analysis on the PSF extension given variation in the learned phase mask coefficients**

**Generalization of DeepFilters to Other Optical NAs to Achieve Leading Extension**

We demonstrate the applicability of DeepFilters for systems with different NA microscope objectives. We are able to optimize EDoFs at the desired NA across each platform while demonstrating minimal mismatch in experimental results. These results encourage the generality of DeepFilters and its potential to be used for platforms across diverse applications.

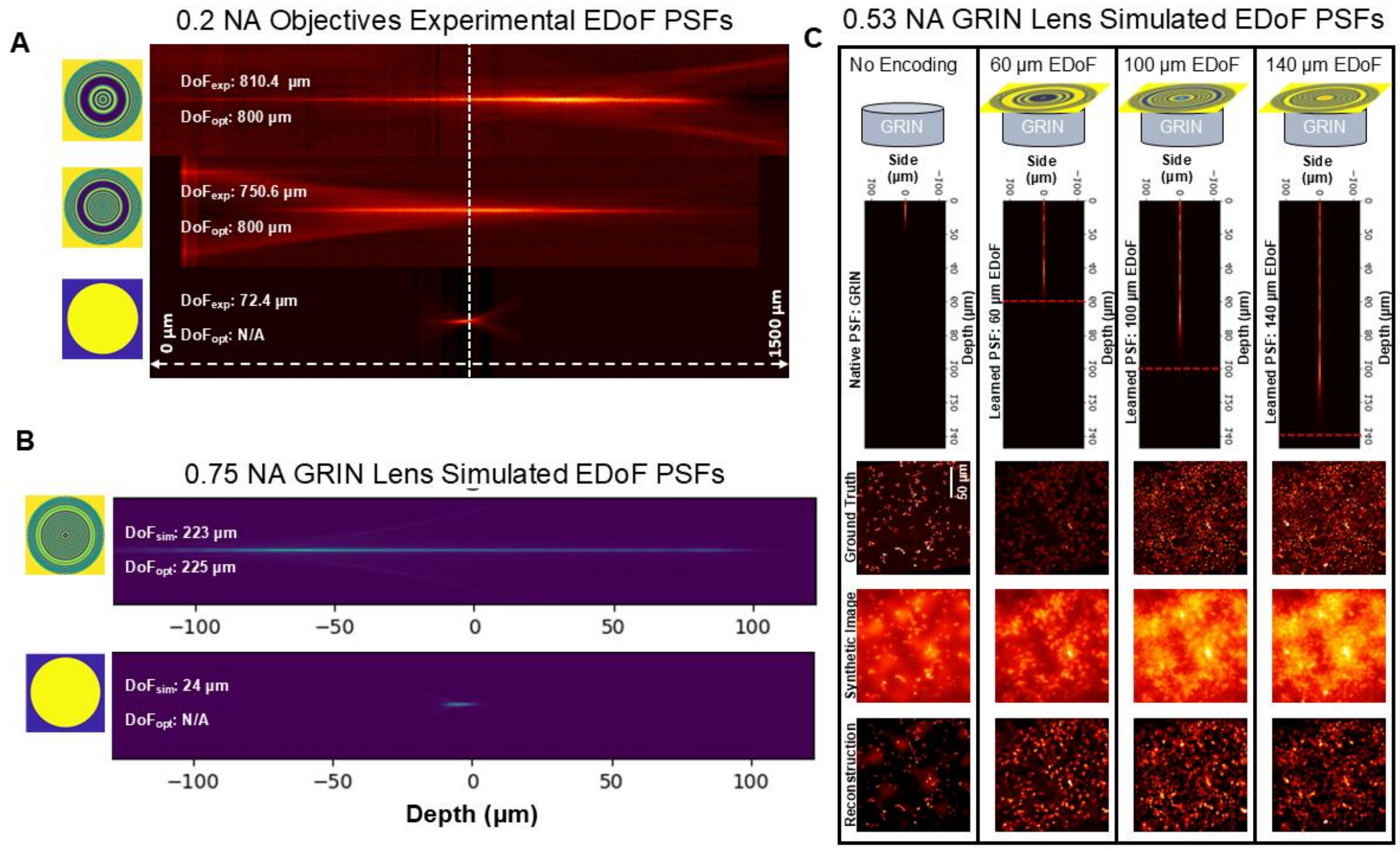

**Fig. S13. Optimization for different NA.** (A) Experimental justification of two optimized 800 µm EDoFs for a 0.2 NA EDoF-Tabletop. (B) Simulated results of an optimized 225 µm EDoF for a 0.75 NA EDoF-Tabletop (C) Simulated results of several EDoFs for a miniaturized zero working distance GRIN EDoF system.

To translate these optimized studies to experiment, we image a tilted sample with 5 µm fluorescent beads with the optimized 0.2 NA EDoF with the corresponding objective on EDoF-Tabletop, as shown in **Fig. S14**.

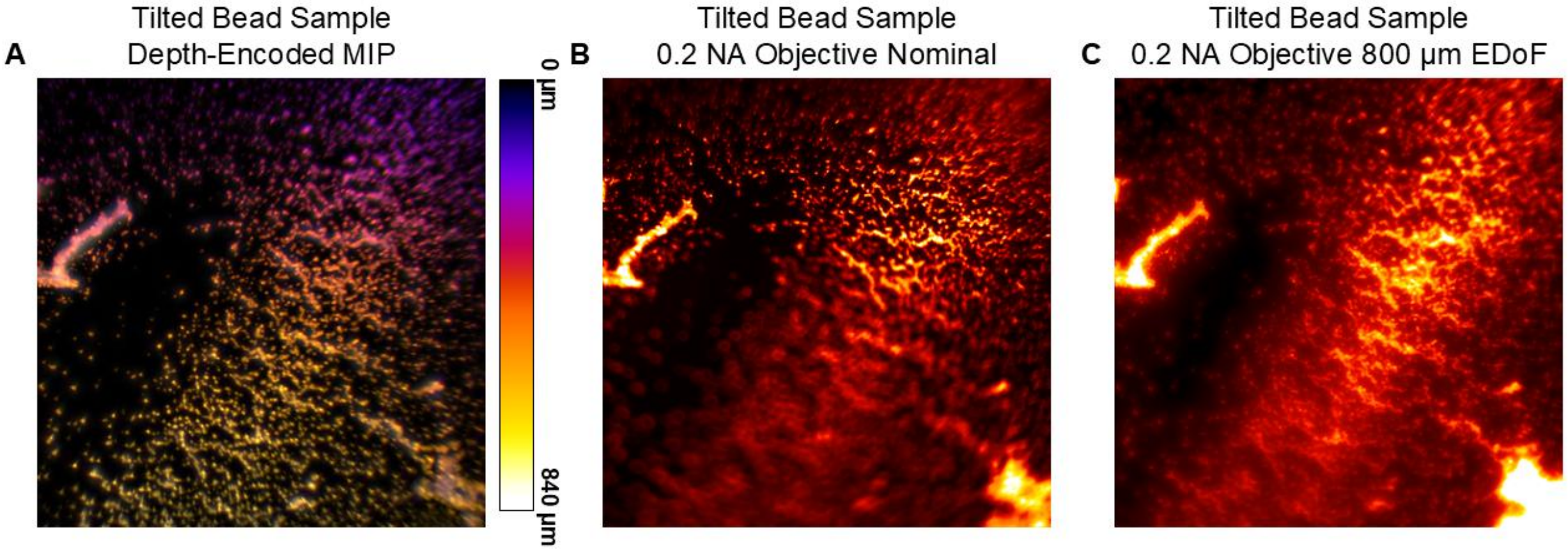

**Fig. S14. Experimental Demonstration of 0.2 NA Optimized EDoF.** (A) Depth-encoded MIP of a tilted 5 µm fluorescent bead sample. (B) Sample imaged under nominal DoF. (C) Sampled imaged under 800 µm optimized EDoF.

**Neuronal Recovery in Fixed Brain Slices of Different Thicknesses**

As demonstrated in section 3.4 in the main text, neuronal signals in fixed mouse brain slices of varying thicknesses can be recovered under different scattering conditions. As increasing the EDOF range does not increase the information recovery without limit due to accumulation of background information, the 140µm mask was deployed across the different samples for consistent recovery around the regime of one scattering length.

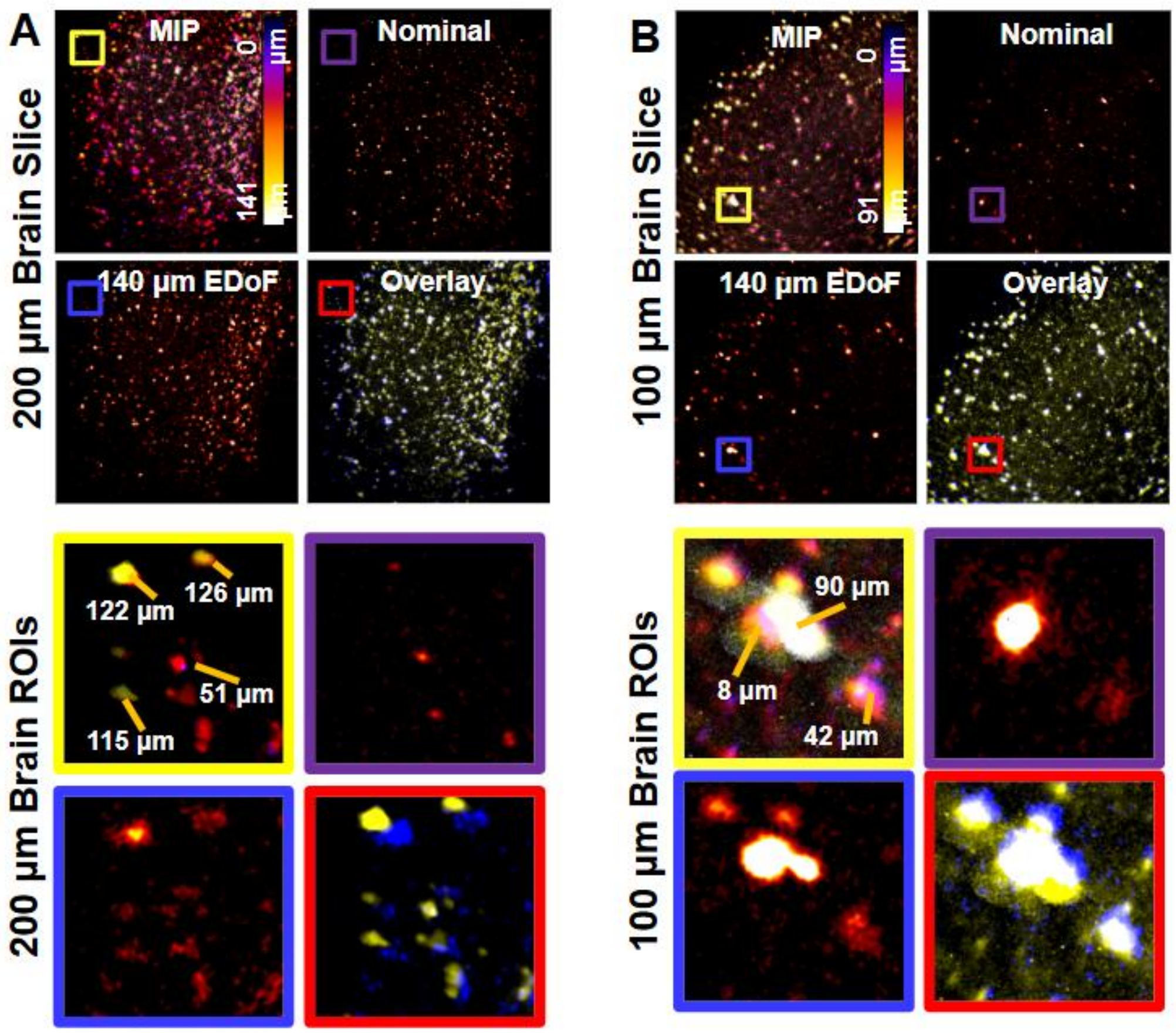

**Fig. S15. Neuronal recovery in fixed mouse brain slices of different thicknesses.** Comparison of depth-encoded MIP, nominal PSF, recovered PSF using 140 µm EDoF mask, and overlay for (A) 100 and (B) 200 µm bran slice with corresponding zoomed in ROIs indicating the deepest recovered neuronal signal.

**Preparation of Proxy Brain Scattering Phantoms**

To test on realistic samples without the need for biological tissue, this work presents a procedure for preparing proxy brain scattering phantoms with realistic properties and morphology. The phantom is fabricated from Formlabs resin with polystyrene beads to set the scattering length and fluorescent beads to act as surrogate neurons, as described in [4], [12]. To inform the mouse brain shape, a meshed mouse brain model is generated and inverted to create a mold, as shown in **Fig. S15A-B.** This mold is scaled to exhibit comparable size and depth to a whole fixed mouse brain, as shown in **Fig. S15C**. The resin is injected into the mold and cured using a Formlabs curing station at 70°C for 1 hour. The fluorescent bead density and phantom thickness is set to produce comparable source density and background to true mouse brains, as shown in **Fig. S15D-F.**

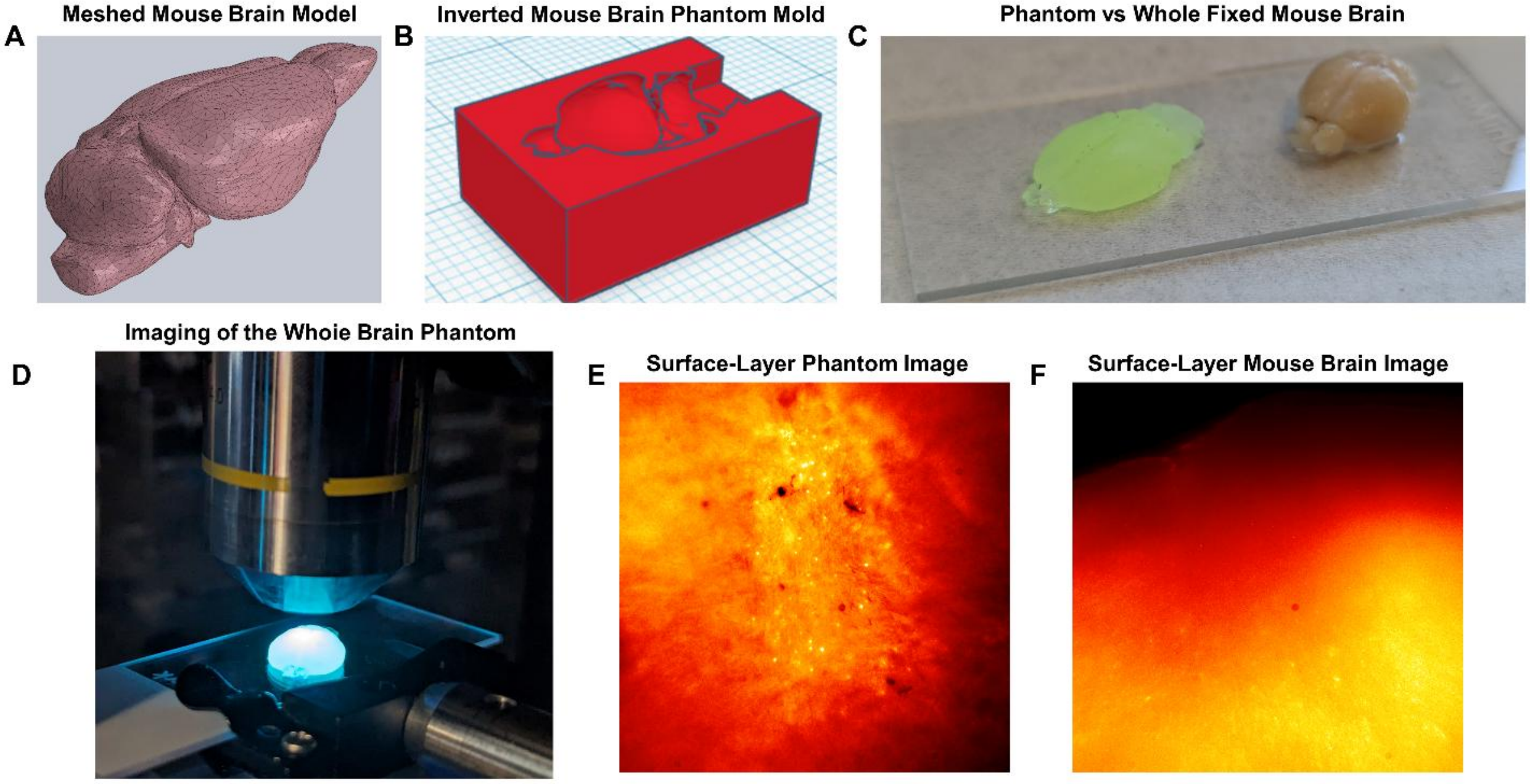

**Fig. S16. Development of Whole Mouse Brain Phantoms.** (A) Meshed mouse brain model used as a reference geometry. (B) Inverted mold used to create phantom. (C) Finalized phantom after curing. (D) Imaging of mouse brain phantom in experiment. (E) Reference image from a finalized mouse brain phantom. (F) Reference image from a thick mouse brain slice.

**Sample Preparation**

The whole brain sample was prepared by injecting a mixture of gelatin (Sigma Aldritch, G2500) in PBS (ThermoFisher Scientific, 10× pH 7.4, 70011069) diluted 1:10 at 10% weight-per-volume for a maximum amount of 10 ml per animal. Next, a heparinized PBS mixture was prepared by adding 600 iu of heparin (ThermoFisher Scientific, H7482) to 30 ml of PBS. Next, 30 mg of fluorescein isothiocyanate (FITC)–albumin (ThermoFisher Scientific, A23015) was added to 1 ml of PBS. Next, the FITC–albumin solution was added to the gelatin solution. A cardiac perfusion was performed using the heparinized saline followed by the FITC–albumin–gelatin. The sample was placed in ice for 15 min. Afterward, the brain was extracted, placed in 4% paraformaldehyde (Electron Microscope Sciences, Diluted 1:8, 15714S) for 6 h, then placed in PBS. To finish the fixation process, the brain in PBS was placed on a horizontal shaker for 3 days.

Wild-caught adult *Lytechinus variegatus* sea urchins were obtained from Pelagic Corp (Sugarloaf Key, FL, USA) or the Duke University Marine Lab (Beaufort, NC, USA). Gametes were collected by injecting adult sea urchins with 0.5 M potassium chloride solution, and embryos were cultured in artificial sea water (ASW) at 23 °C for 48 hours prior to imaging. Embryos were incubated in MitoTracker Green (Invitrogen) diluted to 1:10 in ASW for approximately 5 minutes prior to mounting. Embryos were temporarily immobilized via removal of cilia with 2x ASW and mounted on either poly-L-lysine-treated or untreated slides sealed with Valap (1:1:1 petroleum jelly:lanolin:paraffin).